\documentclass{PoS}

\usepackage{graphicx}
\usepackage{epsfig}
\usepackage{amssymb}

\newcommand{\be}{\begin{equation}}
\newcommand{\ee}{\end{equation}}
\newcommand{\ba}{\begin{eqnarray}}
\newcommand{\ea}{\end{eqnarray}}
\newcommand{\nn}{\nonumber}
\newcommand{\ex}{{\rm e}}
\newcommand{\Tr}  {\mathop{\rm Tr}}

\def\lsi{\raise0.3ex\hbox{$<$\kern-0.75em\raise-1.1ex\hbox{$\sim$}}}
\def\gsi{\raise0.3ex\hbox{$>$\kern-0.75em\raise-1.1ex\hbox{$\sim$}}}
\newcommand{\lsim}{\mathop{\lsi}}
\newcommand{\gsim}{\mathop{\gsi}}

\PoS{PoS(LAT2005)016}

\title{The QCD phase diagram\\ at zero and small baryon density}

\ShortTitle{The QCD phase diagram at zero and small baryon density}

\author{\speaker{Owe Philipsen}\\
        Institut f\"ur Theoretische Physik\\
        Westf\"alische Wilhelms-Universit\"at M\"unster\\
        48149 M\"unster, Germany\\
        E-mail: \email{o.philipsen@uni-muenster.de}}

\abstract{I review recent developments in determining the QCD phase diagram by means of lattice simulations. Since the invention of methods to side-step the sign problem a few years ago, a number of additional variants have been proposed, and progress has been made towards understanding some of the systematics involved. All available techniques agree on the transition temperature 
as a function of density in the regime $\mu_q/T\lsim 1$. There are by now four calculations with signals for a critical point, two of them at  similar parameter values and with consistent results. However, it also emerges that the location of the critical 
point is exceedingly quark mass sensitive. At the same time sizeable finite volume, cut-off and step size effects have been uncovered, demanding additional investigations with exact algorithms on larger and finer lattices before quantitative conclusions can be drawn. Depending on the sign of these corrections, there is ample room for the eventual phase diagram to look as expected or also quite different, with no critical point at all.}

\FullConference{XXIIIrd International Symposium on Lattice Field Theory\\
		 25-30 July 2005\\
		 Trinity College, Dublin, Ireland}

\begin{document}

\section{Introduction}

\subsection{Qualitative expectations \label{sec:qual}}

\begin{figure}
\begin{center}            
{\rotatebox{0}{\scalebox{0.6}{\includegraphics{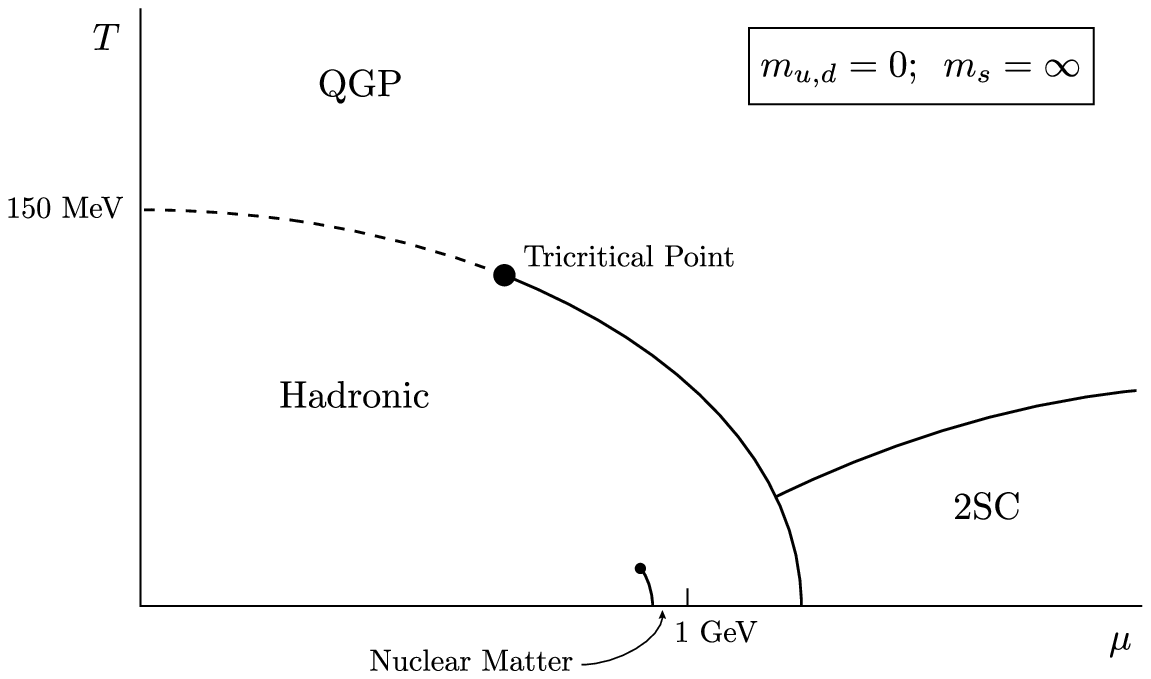}}}}\hspace*{1cm}
{\rotatebox{0}{\scalebox{0.6}{\includegraphics{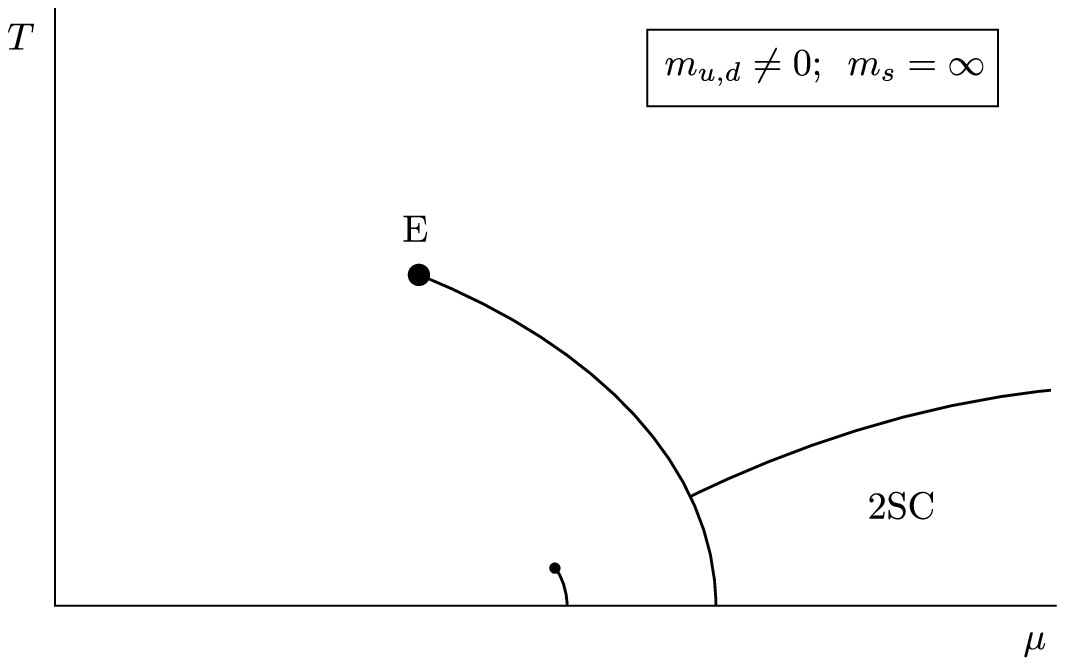}}}}
\end{center}
\caption[]{Qualitative QCD phase diagram for $N_f=2$ according to general expectations. For $N_f=3$
and $m<m_c$, the diagram looks as on the left with the transition being first order all the way, while for $m>m_c$ it looks as on the right. }
\label{qual}
\end{figure}

In the physics communities dealing with QCD at finite temperature and density, there appears
to be little doubt that the $(T,\mu)$ phase diagram qualitatively looks as in Fig.~\ref{qual} (right). 
Given that until recently non-perturbative calculations were impossible for $\mu\neq 0$, and even on the 
temperature axis simulations with dynamical fermions have only become feasible in the last few years,
it seems worthwhile to recall the qualitative arguments that lead to this picture.
Such a diagram represents one set of quark masses. As theorists, we also view the quark masses as parameters and wish to understand the phase diagram of the entire parameter space
$\{m_{u,d},m_s,T,\mu\}$, which should aid us in unveiling the physical situation as well.

Starting point of the argument \cite{wi} is the $N_f=2$ theory with degenerate quark masses. At zero density and in the 
chiral limit, $\mu,m=0$, the chiral condensate represents a true order parameter distinguishing between separate phases, and the symmetry breaking pattern is $SU(2)_V\times SU(2)_A\rightarrow SU(2)_V$. A local order parameter vanishing everywhere in one phase and being non-zero in another
corresponds to a non-analytical function of the parameters of the theory, thus requiring a true phase
transition and excluding an analytical crossover.
{\it If} the corresponding phase transition is second order, {\it then} chiral symmetry implies
that it should be in the universality class of 3d $O(4)$ spin models, a scenario which has been 
very popular among theorists. Note, however, that a first order transition is a logical 
possibility as well. 

For low temperatures and large densities, a number of model calculations (see e.g.~\cite{hal}) 
appear to agree on a
first order transition between nuclear matter and quark matter in a colour-superconducting state, which is expected for asymptotically high density.  
Fig.~\ref{qual} represents only the simplest
picture, other variants have one or several phases
between the hadronic and the superconducting phase, see \cite{hal,murev} for more details. 
The most natural scenario then has the first order line at finite density joining up with the second order line coming from the temperature axis in a tri-critical point.
On the other hand, if the quarks have finite masses, chiral symmetry is explicitly broken and there is no
true order parameter. The chiral condensate still experiences a rapid change of value at the pseudo-critical temperature, but
now it is an analytic crossover. In this case the first order line at finite density 
has to terminate in a critical endpoint.

For three degenerate flavours $N_f=3$, $\mu=0$, the chiral limit exhibits a first order phase transition. 
First order transitions are stable under small variations of the parameters, and thus the first order regime
extends to small masses $m<m_c$, for which it can actually be measured on the lattice.
With increasing quark mass the transition weakens until it ends in a critical point at $m_c$, and for
$m>m_c$ a smooth crossover is observed. For the $(T,\mu)$ phase diagram this implies 
a first oder line connecting the transitions on the axes for $m<m_c$, while for $m>m_c$ the first
order line emanating from the $\mu$-axis again has to terminate in a critical endpoint.

The ``standard scenario'' for the physical case with $N_f=2+1$ quarks is as in Fig.~\ref{qual} (right),
with the critical endpoint moving to larger $\mu_c$ with increasing quark masses,
as may be inferred from a continuity argument.
For $N_f=3$ with $m<m_c$ the phase diagram has a first order line connecting both axes.
Upon sending $m_s\rightarrow \infty$ this picture should continuously evolve into the $N_f=2$ diagram
with a critical endpoint, thus implying $d\mu_c/dm_s>0$.

\begin{figure}[t]
{\rotatebox{0}{\scalebox{0.32}{\includegraphics{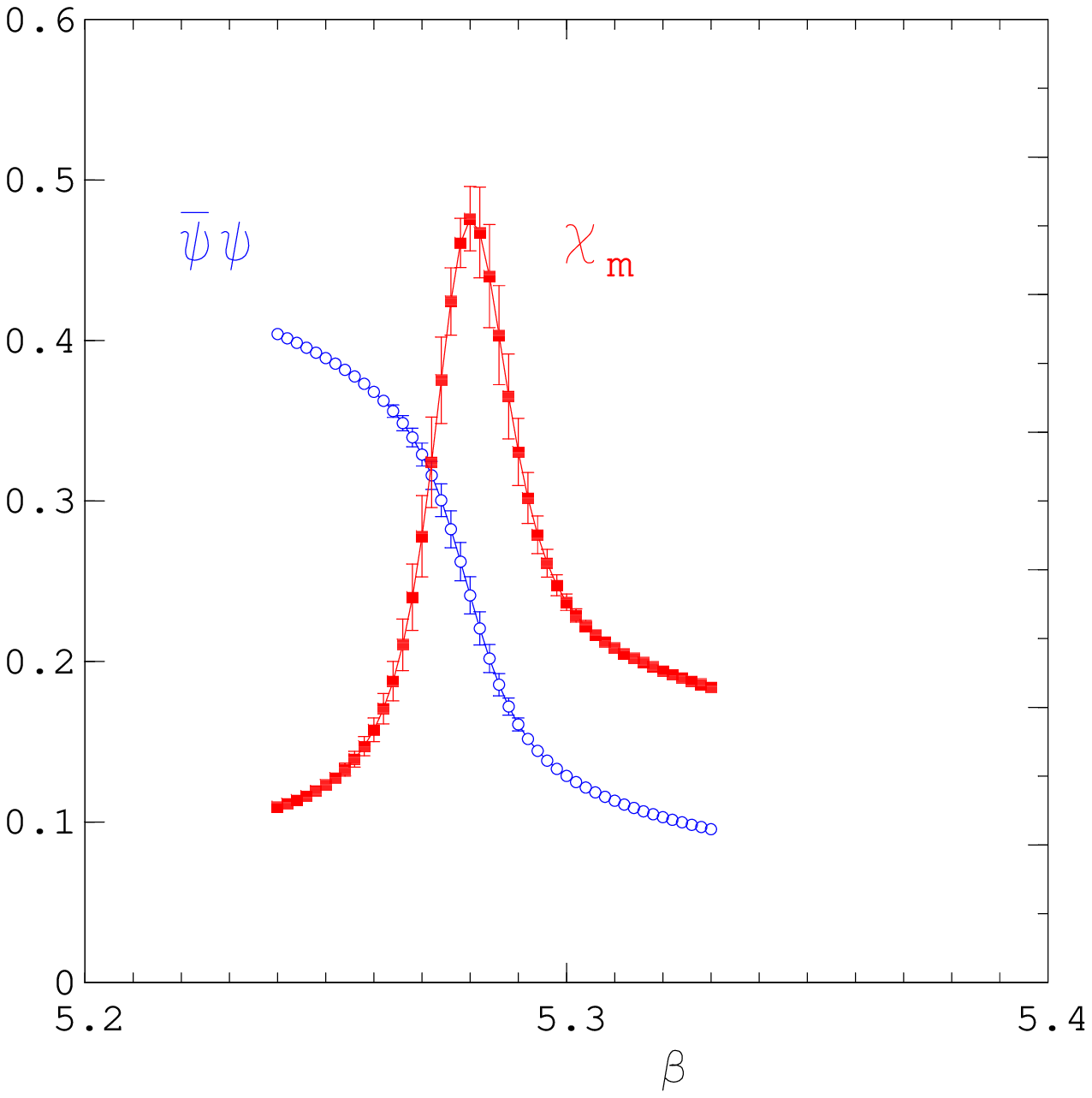}}}}  \hspace*{1.3cm}
{\rotatebox{0}{\scalebox{0.45}{\includegraphics{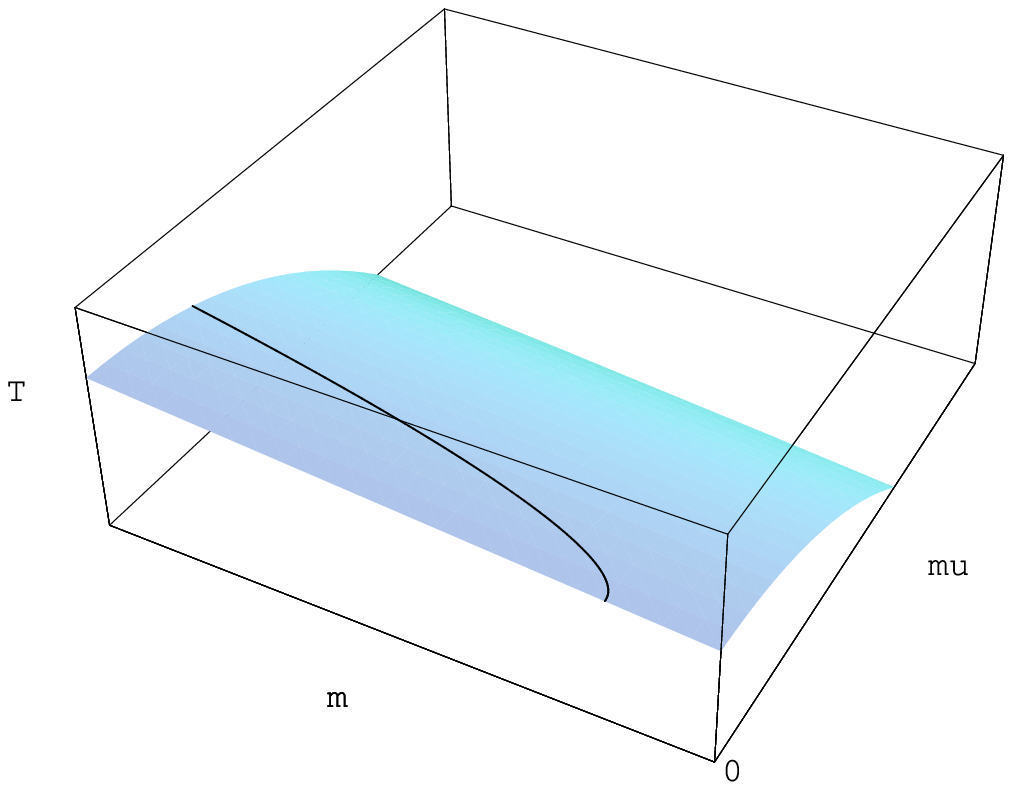}}}} \put(-53,45){1.O.} \hspace*{1cm}   
{\rotatebox{0}{\scalebox{0.45}{\includegraphics{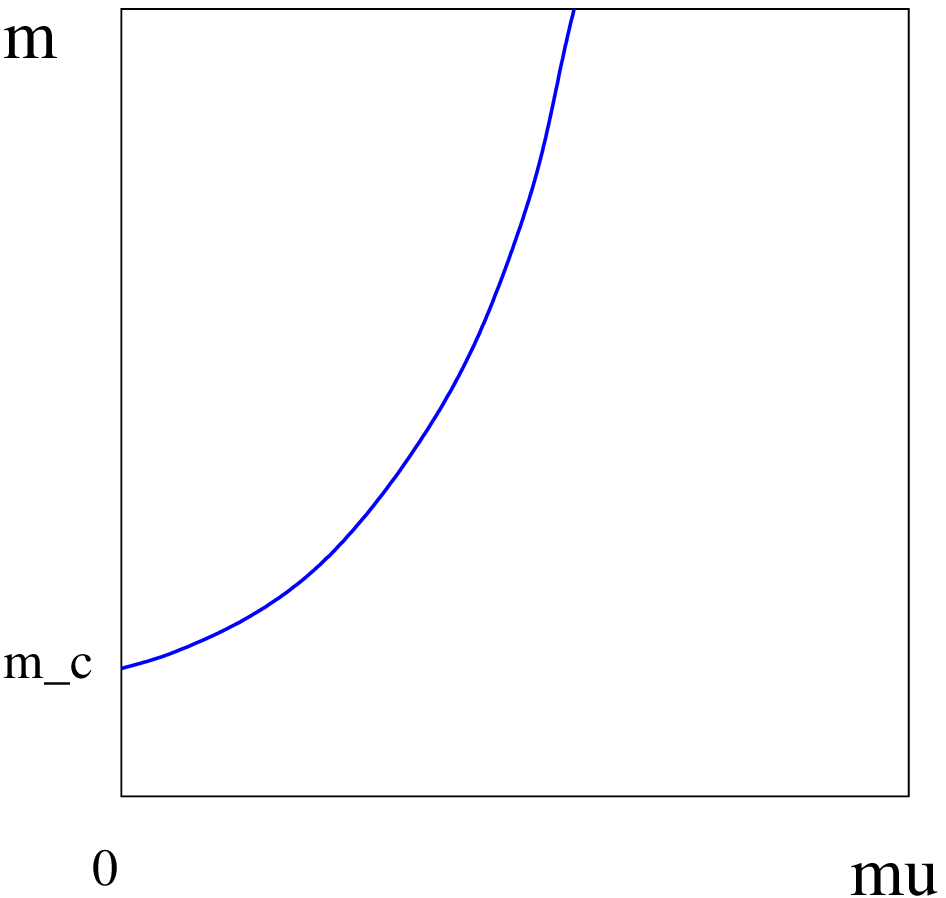}}}}\put(-53,45){1.O.}
\caption[]{Left: Pseudo-critical coupling and temperature defined by the peak of a susceptibility. Middle:  Schematic phase diagram for $N_f=3$. Right: Projection onto the critical surface. }
\label{schem1}
\end{figure}

The full phase diagram of the $N_f=2,3$ theories is in the 3d space $\{m,T,\mu\}$, as in Fig.~\ref{schem1}. 
In order to map it out by simulations the first step is to identify the critical
surface $T_0(\mu,m)$ separating the high and low temperature regions. Since simulations are always on finite volumes, this surface is only pseudo-critical and represents a smooth crossover. It can be 
defined by, e.g., peaks in susceptibilites, cf.~Fig.~\ref{schem1} (left).  
This step is typically rather straightforward.
The much more difficult task is to perform a finite size scaling analysis 
to identify the order of the transition in the infinite volume limit for the different regions of parameter space. For $N_f=3$, such an analysis yields a critical line separating a first order region from a crossover region on the surface 
$T_0(\mu,m)$, Fig.~\ref{schem1} (middle). It is convenient to eliminate the temperature axis from this diagram by projecting onto the 
pseudo-critical surface, i.e.~temperature is always implied to be $T_0(\mu,m)$, Fig.~\ref{schem1} (right).

\begin{figure}[t]
\begin{center}
\vspace*{-6cm}
\leavevmode            
{\rotatebox{0}{\scalebox{0.4}{\includegraphics{phase_diagram_trunc.ps}}}}\hspace*{-1.5cm}
{\rotatebox{0}{\scalebox{0.75}{\includegraphics{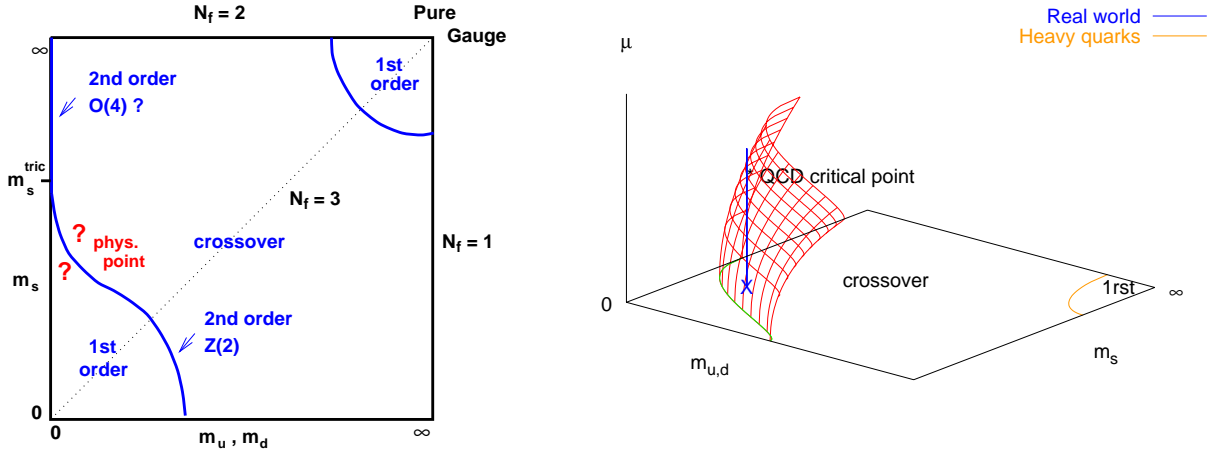}}}}
\end{center}
\caption[]{Left: Schematic phase diagram for $N_f=2+1$ at $\mu=0$. Temperature is implied to be (pseudo-) critical, $T_0( m_{u,d},m_s)$, everywhere. Right: The same with finite quark density.}
\label{schem_2+1}
\end{figure}

This form of a phase diagram is particularly suitable to display the three flavour theory with non-degenerate masses, $N_f=2+1$, including the special cases $N_f=2,3$, as in Fig.~\ref{schem_2+1}. 
Note that because of the difficulties of simulating dynamical fermions, even for $\mu=0$ (left) 
we know very little about the critical lines separating the first order from the crossover regions.
Up to now the only published point that has been calculated to some accuracy with standard staggered fermions is the critical point $m_{u,d}=m_s=m_c$
on the $N_f=3$ diagonal \cite{kls,clm,fp2}, 
which was numerically identified to belong to the universality class of the 
3d Ising model \cite{kls}.  While the statement about the universality class concerns infrared physics and thus is stable against cut-off effects, the location of the critical point in the bare mass diagram 
is very sensitive to renormalisation effects. To date only $N_t=4$ calculations ($a\sim 0.3$ fm) 
have been performed, but simulations with improved actions give values for $m_c$ which are about $\sim 1/4$ of the 
standard action result \cite{kls}. The  critical line for non-degenerate quark masses is being calculated presently, cf.~\cite{fp3} and Section \ref{sec:line}. All available results are consistent with the physical point lying on the crossover side of the boundary. This has also been found in a recent simulation with standard staggered quarks
with a pion to rho mass ratio tuned to its physical value \cite{fk2}. 

When a finite quark number density is switched on, a $\mu$-axis for the chemical potential has to be
added to the diagram, and the critical line separating the first order region from the crossover
region turns into a critical surface, as indicated in Fig.~\ref{schem_2+1} (right). The standard scenario
with $m_c(\mu)$ being an increasing function of the chemical potential then implies that this 
surface bends towards larger quark masses. Consequently, tuning the quark masses to the physical point and switching on a chemical potential, the intersection with the critical surface marks
the critical value $\mu_c$ of the end point, beyond which there is a first order transition.
Thus, a determination of the QCD phase diagram in the full parameter space $\{m_{u,d},m_s,T,\mu\}$ entails mapping out these critical surfaces and understanding how they are joining up in the different limit theories.

\subsection{Lattice QCD at finite temperature and density}
 
Standard Monte Carlo simulations at finite density are made impossible by 
the so-called sign problem of the lattice grand canonical partition function,
\be
Z=\int DU D\bar{\psi} D\psi\,\ex^{-S_g[U]-S_f[U,\psi,\bar{\psi}]}=\int DU\,[\det M(\mu)]^{f}\ex^{-S_g[U]}, \quad S_f=\sum_f \bar{\psi}M \psi.
\ee
For $\mu=0$, the relation $\gamma_5M\gamma_5=M^\dag$ guarantees positivity of the fermion determinant, $\det M\geq 0$, in every gauge background.
 For the gauge group SU(3), the fermion determinant becomes complex as soon as a non-zero quark 
 chemical potential $\mu=\mu_B/3$ is switched on. Thus it cannot be interpreted 
 as  a probability distribution, which rules out standard importance sampling. 
 

The sign problem of QCD is still unsolved to date. Successful solutions of fermion sign problems in
a number of spin models by means of cluster algorithms \cite{wiese} unfortunately do not seem to generalize to QCD. However,
significant progress has been made since 2001 with a number of different approaches
that circumvent the sign problem, rather than solving it. These approaches can be put into three categories:
\begin{itemize}
\item Two-parameter reweighting 
\item Taylor expansion in $\mu/T$
\item Simulations at imaginary $\mu$, either analytically continued to real $\mu$ or Fourier transformed to the  canonical ensemble
\end{itemize}
All these methods have limitations and presently work reliably only for small enough $\mu/T\lsim 1$.
However, the systematics is different between them, thus allowing for meaningful cross checks (for a comparison of early results, see \cite{lp}). 
Another alternative is to study related theories without the sign problem, such as SU(2) QCD and  QCD at finite isospin, the latter being close enough to the case of interest for meaningful comparisons.
The fact that all agree in the determination of the pseudo-critical temperature $T_0(m_i,\mu)$
is one reason for the recent enthusiasm in this field, and gives reason to hope that the order of the transition may be settled in the near future as well.

\section{Massless $N_f=2$ at zero density: O(4) or first order?}

Before delving into the discussion of finite density calculations, let us turn to
the $\mu=0$ behaviour of the theory which played an important role in the derivation of the qualitative phase diagram in Section \ref{sec:qual}. It is a longstanding question whether the phase transition in the chiral limit of the two-flavour theory is indeed second order with O(4) universality or first order. On the lattice, O(4) will effectively look like O(2) as long as there are discretisation effects \cite{o2}.
A lot of work has been done over the years, but no definite conclusion has been reached. Among the more recent work, Wilson fermions appear to see O(4) scaling \cite{wil}, while staggered actions are inconsistent with both O(4) and O(2) \cite{s2}. (The staggered strong coupling limit, however, does display O(2) scaling \cite{c2}). 

A new attempt to tackle this question by means of a finite size scaling analysis with unprecedented lattice sizes was made in \cite{dig}. The work simulates $L^3\times 4$ lattices with 
$L=16-32$,  using the standard staggered action and the hybrid Monte Carlo R-algorithm \cite{ralg}. Several quark masses are studied,  the smallest being $m/T\lsim 0.055$.
In a critical region quantities like, e.g., the specific heat or the chiral susceptibility  
scale universally as
\ba
C_V - C_0 &\simeq & L^{\alpha/\nu} f_c \left(\tau L^{1/\nu}, am\, L^{y_h} \right),
\quad \tau=1-T/T_c\nonumber\\
\chi& \simeq &L^{\gamma/\nu} f_\chi \left(\tau L^{1/\nu}, am\,L^{y_h} \right).
\label{scale}
\ea
Here the non-singular part of the specific heat $C_0$ has been subtracted. 
The values for the exponent $y_h$ are known with some precision and nearly the same for
O(4) and O(2). The authors of \cite{dig} thus fix $y_h$ to this value, and then choose $L$ and
$m$ for a series of simulations such as to keep $(am \,L^{y_h})$ constant. This reduces the two-parameter scaling problem to depend on one remaining variable only, which can be more easily scanned.
The infinite volume limit in this procedure thus corresponds to the chiral limit and allows to check whether the data are consistent with the predicted scaling behaviour.

\begin{figure}[t]
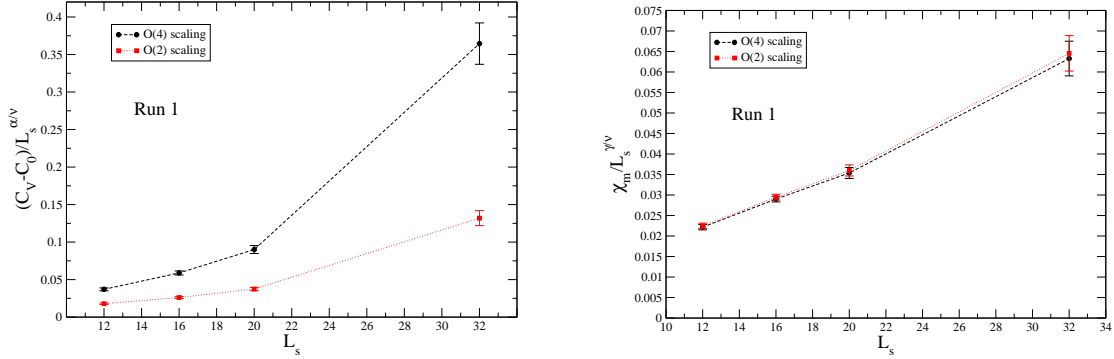

\includegraphics*[width=0.45\textwidth]{Cv_max_Run1.eps}\hspace*{1cm}
\includegraphics*[width=0.45\textwidth]{Chi_max_Run1.eps}
\caption[]{Finite volume scaling behaviour with of specific heat and chiral susceptibility. For O(4) viz.~O(2) behaviour, the data should fall on a horizontal line \cite{dig}.}
\label{dig1}
\end{figure}
\begin{figure}[t]
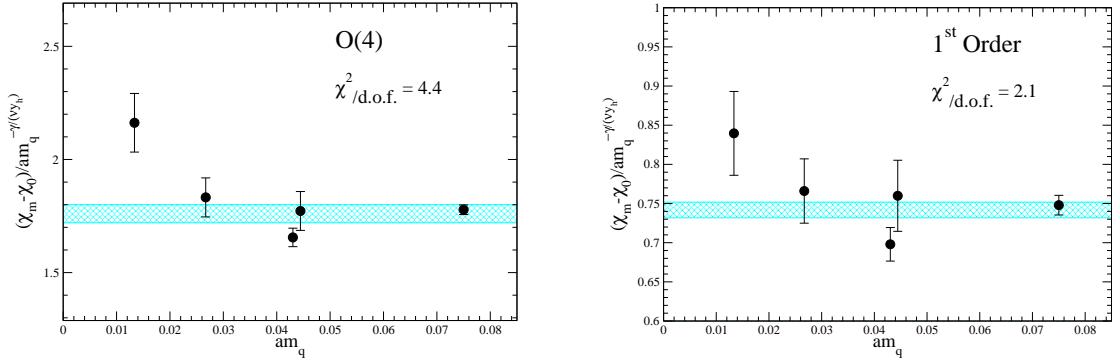

\includegraphics*[width=0.45\textwidth]{Chi_max_O4_2.eps}\hspace*{1cm}
\includegraphics*[width=0.45\textwidth]{Chi_max_1st_2.eps}
\caption[]{Testing the mass scaling of the chiral susceptibility for O(4) and first order behaviour \cite{dig}.}
\label{dig2}
\end{figure}

 Fig.~\ref{dig1} shows simulation results from \cite{dig}. Scaling as in Eq.~(\ref{scale}) would imply the data points to fall on a horizontal line, which is clearly not the case. Alternatively, one may keep the other scaling variable $\tau L^{1/\nu}$ fixed and vary the quark mass. Furthermore, in place of O(4) or O(2) exponents, also consistency with first order exponents can be tried. Some results of this attempt are shown in Fig.~\ref{dig2}. The fit to first order scaling is slightly better, but not very convincing
 either. Moreover, D'Elia et al.~looked in detail at plaquette distributions in the transition region as well as Monte Carlo histories. No signs of a metastability region commensurate with a first order transition could be observed.

Many variations of the analysis are performed in \cite{dig}, with similar outcome. Hence, even 
after formidable computational effort the question of O(4) vs.~first order scaling remains open for the moment. There are several possible explanations. The scaling region away from the chiral limit could be exceedingly small, or discretisation effects could play a large role. This is also suggested by the fact that Wilson and staggered fermions appear to scale differently. The question of cut-off effects will be addressed by the authors of \cite{dig}, who announced an investigation at $N_t=6$ as well.
Another possibility is that exceedingly large volumes might be required to distinguish weakly 
first order and crossover. An example for such behaviour is two-colour QCD in the strong coupling limit.
Numerical results for this theory require lattice sizes $L>128$ before the correct scaling is observed \cite{cj}.  
An important observation is that, 
for both scenarios in Fig.~\ref{dig2}, it is the lowest mass data points which spoil the fits.
This indicates possible systematic errors of the Monte Carlo for very low masses. We shall indeed see in Section \ref{sec:line} that in this regime the R-algorithm has strong step size effects for 
steps of half the quark mass, as chosen for the lowest mass point in \cite{dig}. These effects change the 
apparent order of the phase transition. Thus, for any future investigation an exact algorithm is necessary.
Meanwhile, we should keep an open mind to the possibility of a first order transition in the chiral limit. In this case the phase diagram Fig.~\ref{qual} would be as for $N_f=3$ with a very small critical quark mass $m_c$. There would be a first order line all the way for $m<m_c$, or a first order line with an endpoint,  Fig.~\ref{qual} (right), for $m>m_c$.

\section{Finite density phase diagram from two parameter reweighting}
 
%

\begin{figure}[t]    
\vspace*{-2cm}
{\rotatebox{0}{\scalebox{0.4}{\includegraphics{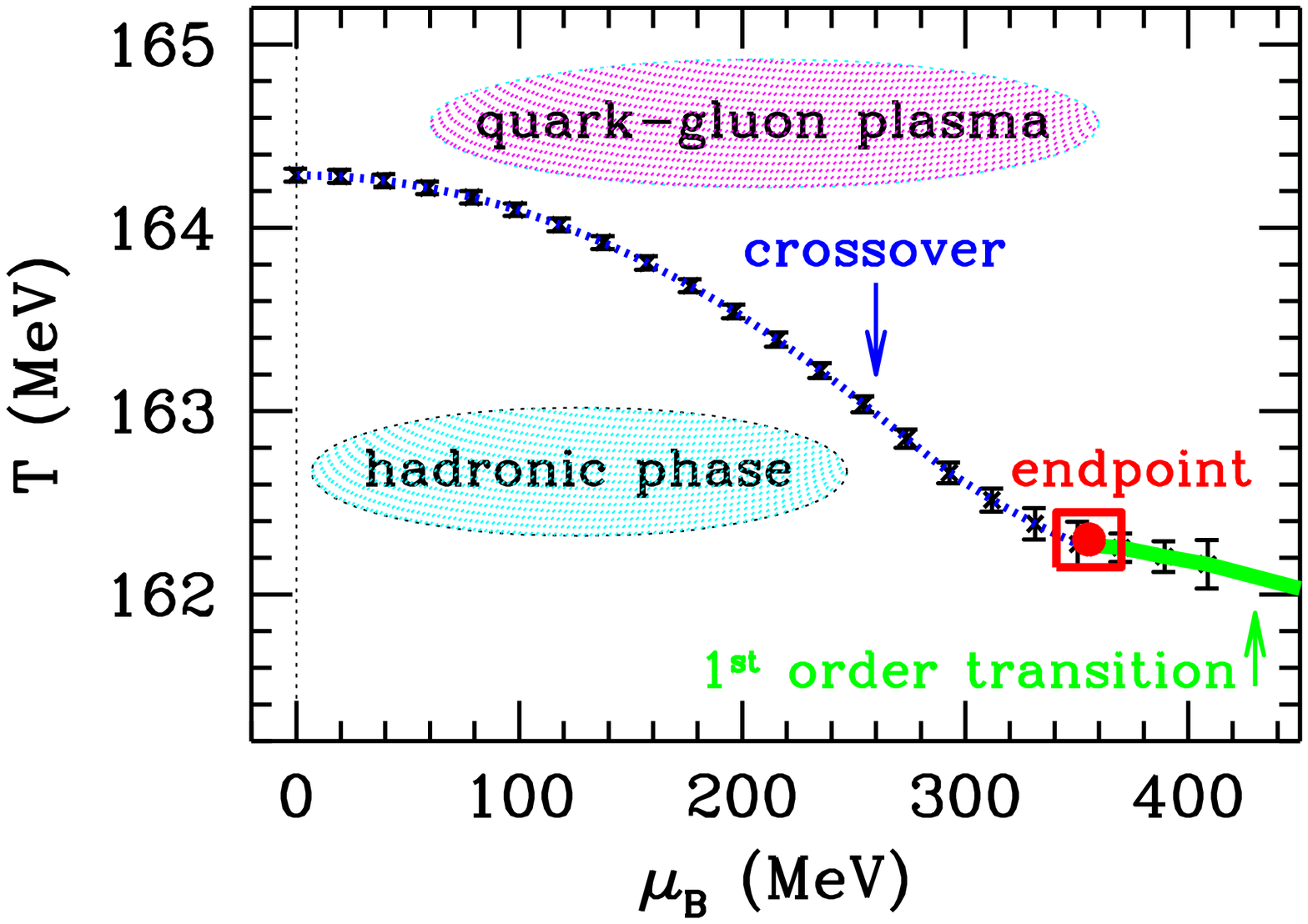}}}}\hspace*{1cm}
{\rotatebox{0}{\scalebox{0.75}{\includegraphics{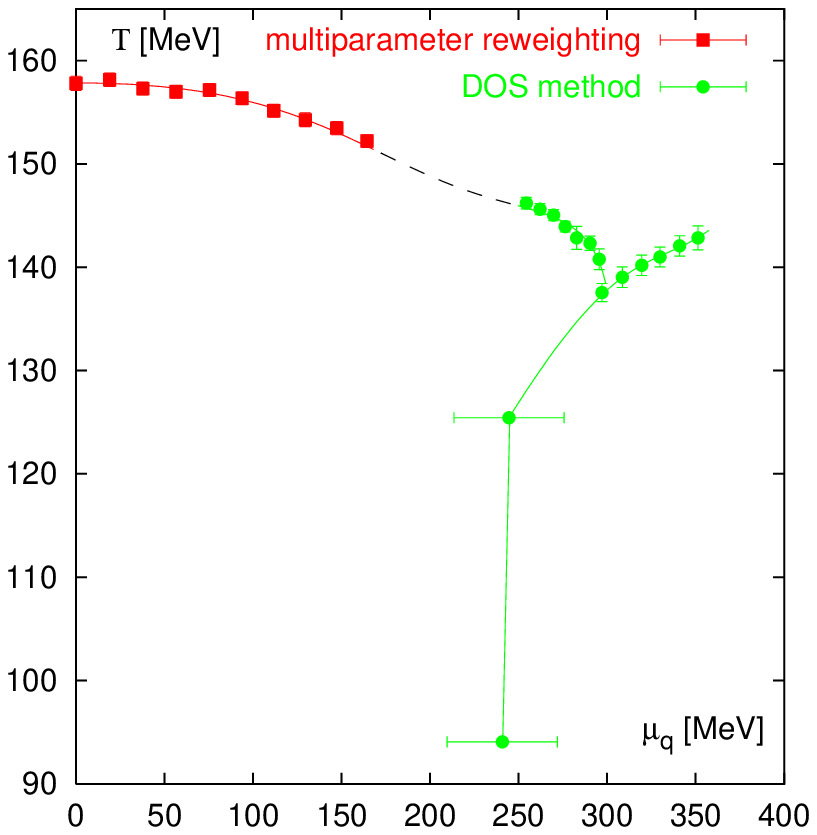}}}}
\caption[]{Left: The phase diagram for physical quark masses as predicted by the two parameter reweighting method \cite{fk2}. Right: High density results from the density of states method, indicating a triple point \cite{cs}. }
\label{fk}
\end{figure}

Significant progress enabling finite density simulations was made a few years ago, by a generalisation of the Glasgow method \cite{gla} to reweighting in two parameters \cite{fk0}.  
The partition function is rewritten identically as
\be 
Z=\left \langle \frac{\ex^{-S_g(\beta)}\det(M(\mu))}
{\ex^{-S_g(\beta_0)}\det(M(\mu=0))}\right\rangle_{\mu=0,\beta_0},
\ee
where the ensemble average is now generated at $\mu=0$ and a lattice gauge coupling 
$\beta_0$, while a reweighting factor takes us to the values
$\mu,\beta$ of interest. 
The original Glasgow method reweighted in $\mu$ only and was suffering from the
overlap problem: while the reweighting formula is exact, its Monte Carlo evaluation is not. The integral
gets approximated by a finite number of the most dominant configurations, which are different for the reweighted and the original ensemble, and this difference grows with $\mu$.
When calculating critical behavior at some $\mu$, one-parameter reweighting uses a non-critical ensemble at $\mu=0$, thus missing important dynamics.
By contrast, two-parameter reweighting proceeds along the pseudo-critical line of the phase change, thus always working with an ensemble that probes both phases. This approach produced the first finite density phase diagram from the lattice, obtained for light quarks corresponding to $m_\pi\sim 300$ MeV  \cite{fk1}. A Lee-Yang zero analysis \cite{lyz} was employed in order to find the change from crossover behaviour at $\mu=0$ to a first order transition for $\mu>\mu_c$.  
A later simulation at physical quark masses puts the critical point 
at $\mu_B^c\sim 360$ MeV \cite{fk2}, Fig.~\ref{fk} (left). In this work $L^3\times 4$ lattices with $L=6-12$
were used, working with the standard staggered fermion action and using the R-algorithm. Quark masses were tuned to $m_{u,d}/T_0\approx 0.037, m_s/T_0\approx 1$, corresponding to the mass ratios $m_{\pi}/m_{\rho}\approx 0.19, m_{\pi}/m_K\approx 0.27$, which are close to their physical values.

A difficulty in this approach is that the determinant
needs to be evaluated exactly.  Because of the sign problem the reweighting factor is exponentially suppressed with volume and chemical potential, thus limiting the applicability to moderate values of those parameters. This point will be discussed in more detail later.
Moreover, since the statistical fluctuations are those of the simulated ensemble instead of the physical one,  it remains difficult to obtain reliable error estimates. For a proposed procedure see \cite{fk3}. 

In present work in progress two-parameter reweighting is combined with the density of states method \cite{dos}, in order to extend the applicability of reweighting to larger values of $\mu/T$ and thus to lower $\mu$. 
First interesting results,  with indications for a possible triple point, are  shown in Fig.~\ref{fk} and presented in more detail in these proceedings \cite{cs}. 

\section{Finite density by Taylor expansion}

Another method to gain information about non-zero $\mu$ is to compute the coefficients of a Taylor series expansion of observables in powers of $\mu/T$. 
Early attempts have looked at susceptibilities and the response of screening masses to chemical potential \cite{milc,taro,hlp,gg}. More recently it has also been used to gain information on the phase transition and its nature itself \cite{bisw1}-\cite{ggpd}.
This idea exploits the fact that on finite volumes there are no non-analytic transitions, and hence the partition function $Z(m>0,\mu,T)$ is an analytic function of the parameters of the theory.
For small enough $\mu/T$ one may then hope to get away with only a few terms, whose coefficients are calculated at $\mu=0$. 
Moreover, CP symmetry of the QCD action translates into  a reflection symmetry of the partition function, $Z(\mu)=Z(-\mu)$, such that real physical observables have series expansions in $(\mu/T)^2$. Thus, in particular the pressure density can be expressed as an even power series, 
\be
 p(T,\mu)=-\,\frac{F}{V} 
= \left(\frac TV\right)\log Z(T,\mu),\quad
{p\over T^4}=
\sum_{n=0}^\infty c_{2n}(T) \left({\mu\over T}\right)^{2n}.
\label{press}
\ee
Since only even terms appear, the coefficients are equivalent to generalised quark number susceptibilities at $\mu=0$, and hence measureable with standard simulation techniques.
For high enough temperatures $T>T_0$, the scale of the finite temperature problem is set by the 
Matsubara mode $\sim \pi T$, and one would expect coefficients of order one for an expansion in 
the `natural' parameter $\mu/(\pi T)$ \cite{fp2}. We shall see later that this is borne out by simulation results.

Since all the $\mu$-dependence in the partition function is sitting in the fermion determinant, it is derivatives of the quark matrix that need to be computed,
\be
\frac{\partial \ln \det M}{\partial \mu}  ={\rm tr} \left( M^{-1} \frac{\partial M}{\partial \mu} \right),\quad
\frac{\partial {\rm tr} M^{-1}}{\partial \mu} =
- {\rm tr} \left( M^{-1} \frac{\partial M}{\partial \mu}
 M^{-1} \right), \quad \mbox{etc.},
 \ee
which can be iterated for higher orders. These expressions become increasingly complex and methods
to automatize their generation have been devised \cite{ggpd}.
Note that one now is dealing with traces of composite local operators, which greatly facilitates the numerical evaluation in a simulation compared to a computation of the full determinant.
The numerical estimate of these expressions proceeds by the random noise method, with typically
$O(10)-O(100)$ Gaussian noise vectors.  

If one is interested in phase transitions, finite volume scaling towards the thermodynamic  limit has to be considered. True phase transitions will emerge as non-analyticities in the pressure, which is not the case for analytic crossover behaviour. 
Given that in the two flavour theory with finite masses and for $\mu=0$ the deconfinement transition  is an analytic crossover, one may expand about $\mu=0$ and then look for the emergence of a finite radius of convergence as the volume increases. 
The radius of convergence of a power series gives the distance between the expansion point and the nearest singularity, and may be extracted from the high order behaviour of the series. Two possible 
definitions are
\be
\rho,r = \lim_{n\rightarrow\infty}\rho_n,r_n\qquad \mbox{with}\quad
    \rho_n=\left|\frac{c_0}{c_{2n}}\right|^{1/2n},
     \qquad
   r_n=\left|\frac{c_{2n}}{c_{2n+2}}\right|^{1/2}.
\label{rad}
\ee
General theorems ensure that if the limit exists and asymptotically all coefficients of the series are positive, then there is a singularity on the real axis. 
More details as well as previous applications to strong coupling expansions 
in various spin models can be found in \cite{series}.
In the series for the pressure such a singularity would correspond to the critical point in the $(\mu,T)$-plane.
 
The study of finite size scaling of a Taylor series presents a formidable technical task. Since the coefficients are generalised susceptibilities, each of them exhibits non-trivial finite size scaling. The scaling of the individual coefficients, evaluated at $\mu=0$, has to combine to the correct scaling of the finite density pressure given by the sum, thus requiring delicate cancellations between the individual contributions in the large volume limit. Classifications of the behaviour of various generalised susceptibilities are given in \cite{gg2}. 

\subsection{Quark number susceptibility to order $\mu^6$ for $N_f=2$}

New results from this approach were reported this year by Gavai and Gupta \cite{ggpd}. They perfomed simulations on $L^3\times 4$ lattices with $L=8-24$, using the standard staggered action and the R-algorithm. The quark mass was fixed in physical units to $m/T_0=0.1$. The aim of the simulations was to bracket the critical point by computing the Taylor coefficients of the quark number susceptibility up to sixth order (i.e.~8th order for the pressure) for various temperatures in the range $T/T_0=0.75-2.15$, and extrapolate to finite $\mu$.
This was done for different lattice volumes in order to get an estimate of finite voulme effects.

\begin{figure}[t]
\includegraphics*[width=0.45\textwidth]{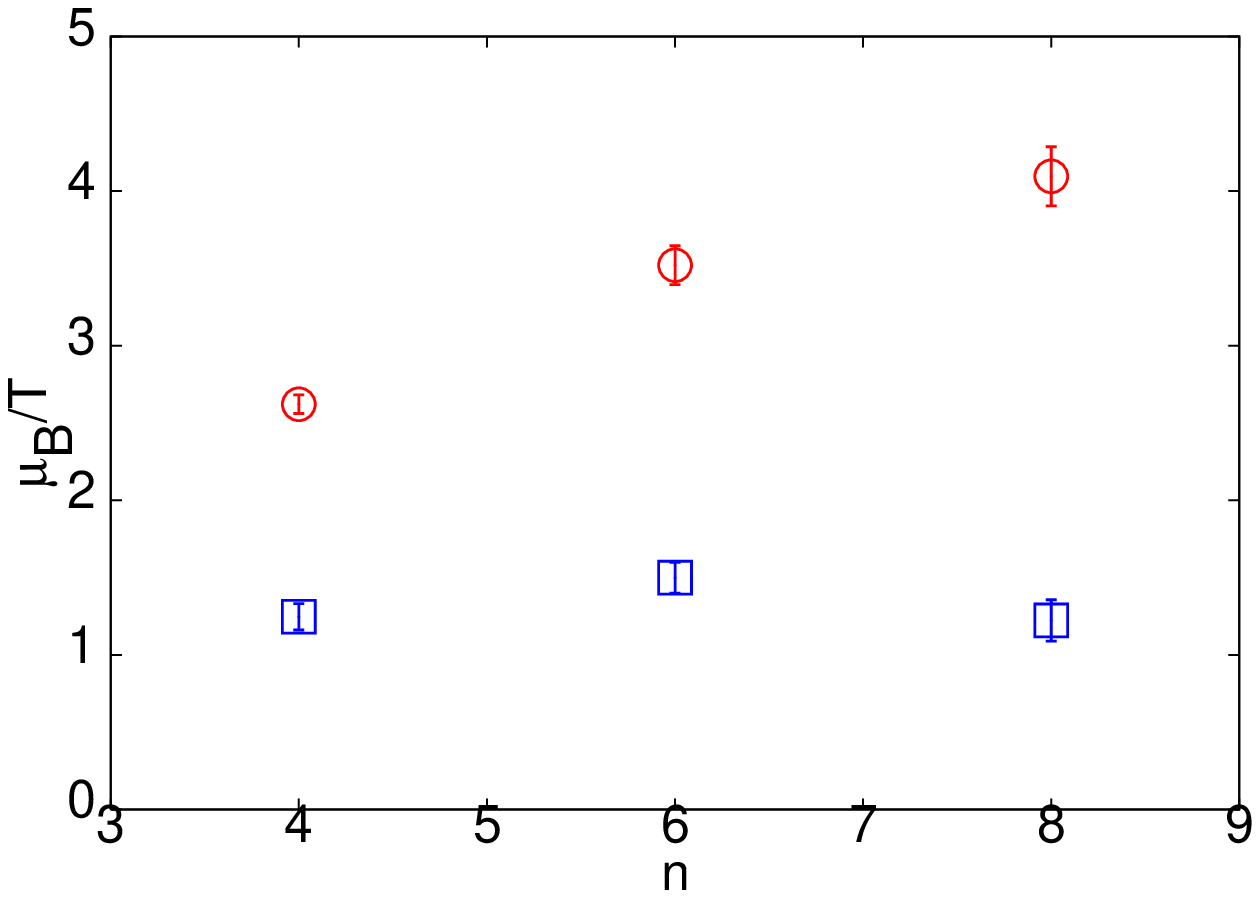}\hspace*{1cm}
\includegraphics*[width=0.45\textwidth]{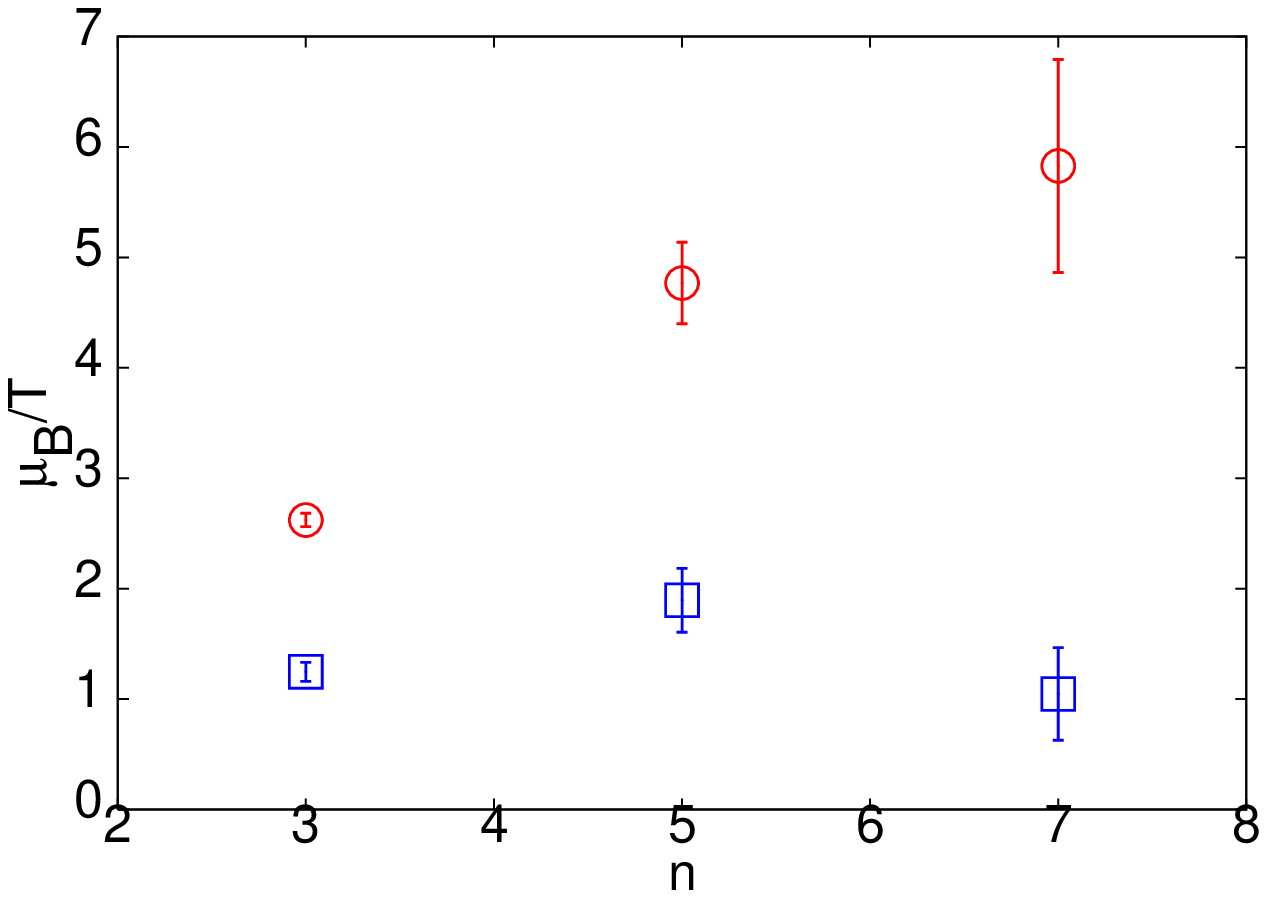}
\caption[]{Estimators of the radius of convergence, Eq.~(\ref{rad}), at $T/T_0=0.95$. $\rho_n$ (left) and $r_n$ (right) vs.~order $n$ on $8^3$ (red circles) and $24^3$ (blue squares) \cite{ggpd}.}
\label{ggfig}
\end{figure}

The results for the convergence radius Eq.~(\ref{rad}) are shown in Fig.~\ref{ggfig}. A rather strong volume dependence is apparent. While for the smaller $8^3$ lattice the estimators $\rho_n,r_n$ do not seem to converge to a finite radius of convergence, the results on the larger $24^3$ lattice are consistent with settling at a limiting value. The boundary between the two behaviours was observed to occur at $Lm_\pi\approx 5-6$ or $L\approx 16-18$, with larger volumes tending to a smaller radius of convergence. It is reassuring that the numerical values for $\rho_n$ and $r_n$ are consistent with each other.
Taking the large volume result at face value and extrapolating to all orders the estimate 
for the location of the critical point is $\mu^c_B/T=1.1\pm 0.2$ at  $T/T_0=0.95$, or 
$\mu_B^c/T_c=1.1\pm 0.2$ \cite{ggpd}.

\subsection{The pressure to order $\mu^6$ for $N_f=2$}

Another investigation of the finite density phase diagram of the two-flavour theory was made by
the Bielefeld-Swansea collaboration, also using the Taylor expansion of the pressure. 
This group works with a $16^3\times4$ lattice with p4-improved staggered fermions and a Symanzik-improved Wilson action, simulating with the R-algorithm, 
the quark mass is set to $m/T_0\approx0.4$.  The calculation to order $\mu^4$ was performed in \cite{bisw2} while new results on $\mu^6$ are presented in \cite{bisw3}. The last work also contains detailed discussions of two interesting analytic calculations to compare with,
namely the pressure in high temperature perturbation theory \cite{vuo}, which is going to hold at asymptotically high temperatures, as well as the hadron resonance gas model, which gives a rather good description of the pressure in the confined phase \cite{krt}.

\begin{figure}[t]
\begin{center}
\includegraphics*[width=0.45\textwidth]{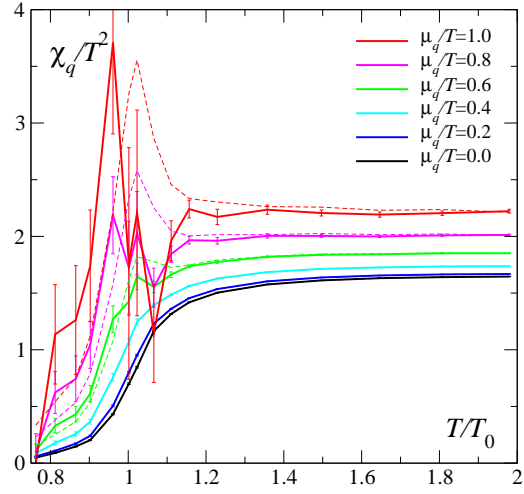}
\end{center}
\caption[]{Quark number susceptibility computed through $O(\mu^4)$ (dashed lines) and $O(\mu^6)$ (solid lines) \cite{bisw3}.}
\label{susc}
\end{figure}

In agreement with \cite{ggpd} and qualitative expectations, their detailed results for the coefficients in the pressure series satisfy $c_6\ll c_4 \ll c_2$ for $T>T_0$, i.e.~one would have coefficients of order one for an expansion in $(\mu/\pi T)$. An impression of the convergence of the series can be obtained by looking at the quark number susceptibility calculated to consecutive orders, as shown in Fig.~\ref{susc}.
For $T\lsim 1.2 T_0$, the series seems to converge rapidly and the $\mu^6$-result is compatible with the one through order $\mu^4$. Around the transition temperature $T_0$, the $\mu^4$-results show a peak emerging with growing $\mu/T_0$, which in \cite{bisw2}  was interpreted as evidence for
a critical point. However, the $\mu^6$ contribution suggests that in this region results do not yet converge,  and the structure is hence not a significant feature of the full pressure.

\begin{figure}[t]
\includegraphics*[width=0.45\textwidth]{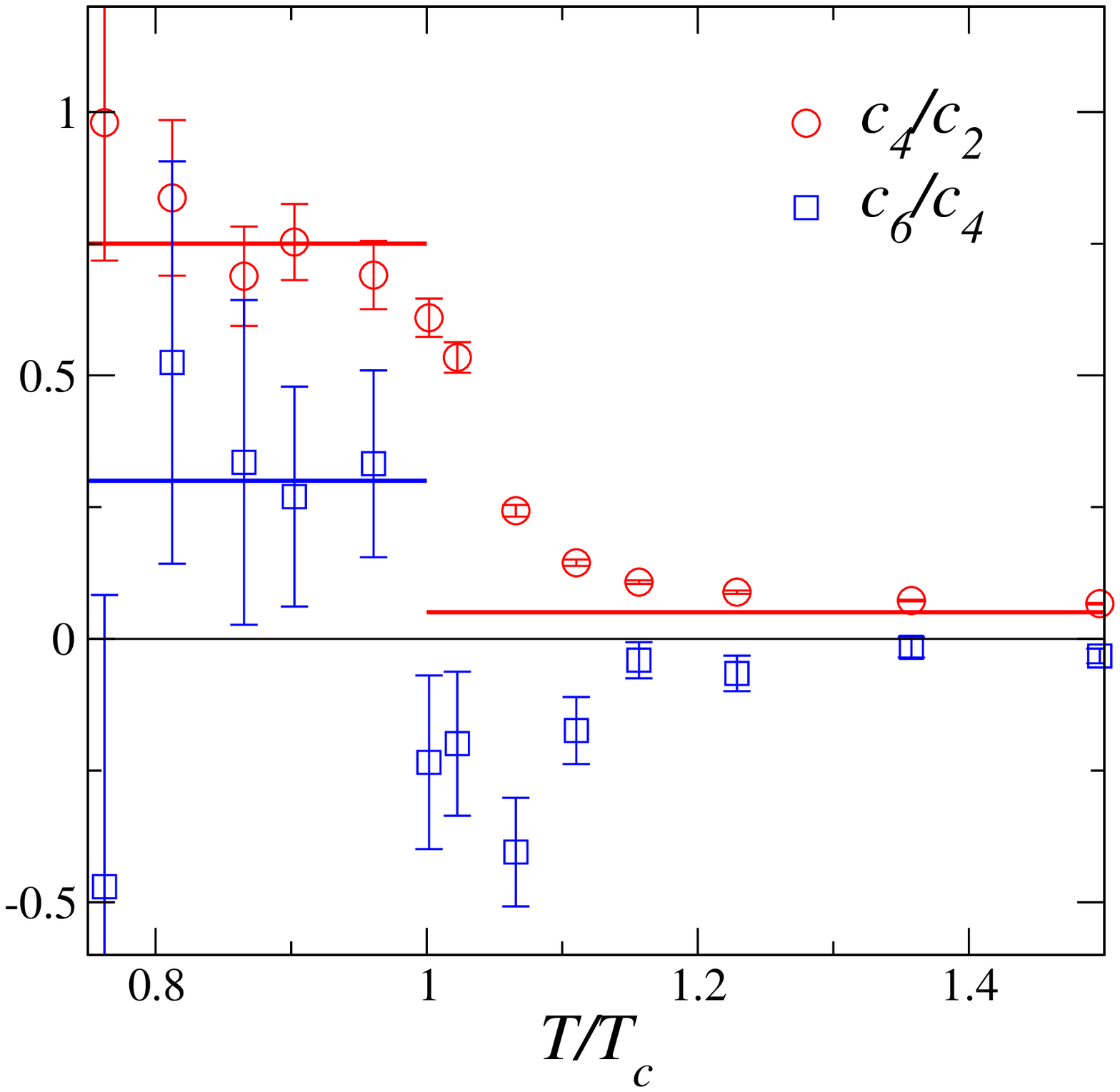}\hspace*{1cm}
\includegraphics*[width=0.45\textwidth]{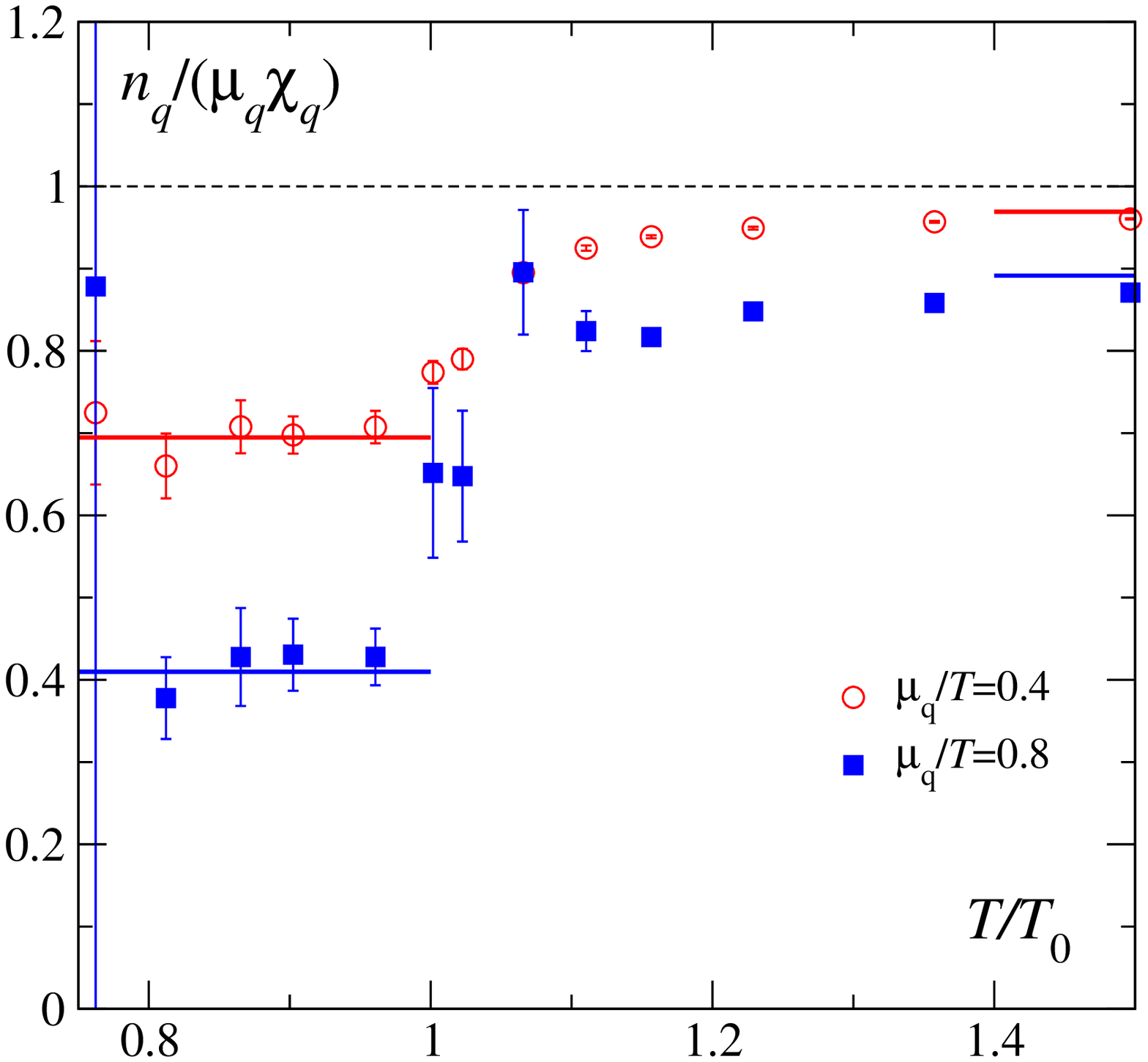}
\caption[]{Left: Ratios of the expansion coefficients, Eq.~(\ref{press}). Right: Quark number density divided by the susceptibility. This observable goes to zero at a critical point. Solid lines in both plots represent predictions from the hadron resonance gas $(T<T_0)$ and the Stefan-Boltzmann ideal gas $(T>T_0)$. \cite{bisw3}.}
\label{srad}
\end{figure}

Another analysis of the data in \cite{bisw3} is devoted to a study of the convergence radius. Fig.~\ref{srad} (left) shows the ratio of the Taylor coefficients at consecutive orders. In the relevant region $T\lsim T_0$, the data do not seem to settle on a limit value. 
More strikingly, the data appear to fall right onto the solid lines
marking the prediction for those ratios from the hadron resonance gas model, on both sides of the transition. The hadron resonance gas model does predict a Hagedorn-like deconfinement transition, but as a smooth crossover rather than a real phase transition. Thus the data do not give any indication of a critical point.  
This is corroborated by Fig.~\ref{srad} (right), showing the quark number density normalised on the susceptibilty, $n_q/\chi_q=\frac{\partial p}{\partial n_q}$. This quantity is related to the compressibility in the plasma, and should go to zero at a second order phase transition point. 

Thus the conclusion in \cite{bisw3} is that there is no evidence for a citical point from these data. This conclusion is not in conflict with the results from \cite{ggpd} discussed in the previous section, since the simulations were done at a much larger quark mass, for which one would expect a critical point to be at larger values of $\mu$. Moreover, a different action was used, making a direct comparison difficult.
However, the conclusion is different from the earlier one by the same group based on $\mu^4$ results \cite{bisw2},
which were interpreted as showing evidence for a critical point.
This highlights the need for a careful examination of as many terms in the series as possible, before results are conclusive. Indeed, previous experience with inferring phase structures from convergence properties of strong coupling series in spin models \cite{series} shows that this is
very much an ``experimental science''. Between 10-20 terms are known in some of these expansions,  and while for some models stunningly good predictions about non-analytic behaviour are obtained, others still fail even at this high order. 

\section{QCD at finite isospin density}

Another way to learn about QCD at finite density is by taking recourse to theories without
a sign problem, which are sufficiently close enough to the physical situation of interest. 
I shall not go into the long list of activities along those lines, but concentrate on
QCD at finite isospin density with chemical potential $\mu_I$ \cite{ss}. For small enough $\mu_I/T$ it can be argued that this theory should agree quantitatively with that at small $\mu/T$, and recent lattice simulations support this picture \cite{ks1,ks2}. 

\begin{figure}[t]
\begin{center}
\includegraphics*[width=0.45\textwidth]{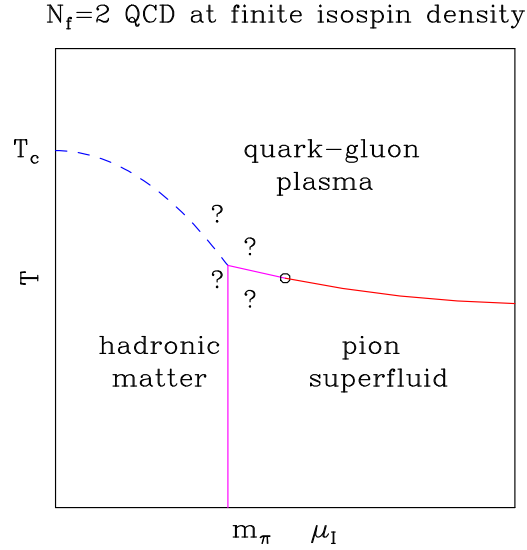}
\caption[]{Schematic phase diagram for QCD at finite isosipin density.}
\label{iso}
\end{center}
\end{figure}

QCD at finite isospin density is obtained
from two-flavour QCD by assigning opposite chemical potentials to the quark flavours,
$\mu_u=-\mu_d=\mu$, leading to $\mu_I\equiv (\mu_u-\mu_d)=2\mu$. This results in cancelling the 
phase of the determinant, so that the partition function now contains only its modulus and thus has
real positive measure, which can be simulated without problems,
\be
Z=\int DU\,|\det M(\mu)|^{N_f}\ex^{-S_g[U]}.
\ee
A schematic phase diagram of the theory is shown in Fig.~\ref{iso}. On the lower right it features a pion 
superfluid phase, due to pion condensation $\langle\pi^-\rangle\neq 0$ when $|\mu_I|>m_\pi$.  
Note that in nature one cannot have a system with $\mu_I\neq 0$ and $\mu_B=0$, since the weak interactions do not conserve isospin. The interest in this theory is because of its formal 
relation to QCD at finite baryon density. It is also in this formal sense that the concept is generalised to 
$N_f=3$.  
Indeed one would expect that for $\mu_I$ sufficiently small it should recover the physics at small baryon density, and hence the dashed transition line in Fig.~\ref{iso} should be approximately the same as in the theory with baryon density.
The argument goes as follows \cite{ks1}. 
An expectation value evaluated at finite baryon density can be rewritten as an expectation value evaluated at finite isospin density  by means of the reweighting formula,
\be
\langle O \rangle_\mu={\langle e^{i\theta} O \rangle_{\mu_I=2\mu}
                   \over      \langle e^{i\theta} \rangle_{\mu_I=2\mu}}.
\ee
Now consider probing the deconfinement transition with a gluonic observable $O$, e.g.~ the plaquette
susceptibility showing a peak. As long as $\langle \cos\theta\rangle_{\mu_I}\sim 1$, the observable
$O'=e^{i\theta} O $ will signal the same transition as $O$. But in this regime one may as well neglect 
the phase altogether, in which case one probes the transition at finite isospin density.
Based on this argument, one expects the transition lines in the two theories to be close to each other as long as reweighting works, i.e.~for $\mu/T_0\lsim 1$. (In an expansion about $\theta=0$, the difference should be of the order $\sim \langle \theta^2\rangle$).   
This expectation was numerically verified for the pseudo-critical surface $T_0(m,\mu)$ in the $N_f=2,3$ 
theories. In the two-flavour theory, the Bielefeld-Swansea collaboration performed a Taylor expansion both in baryon and isospin chemical potential, and the resulting pseudo-critical lines were found to quantitatively agree \cite{bisw2}. Similarly, quantitative agreement was found between the pseudo-critical lines determined from finite $\mu_I$ \cite{ks1,ks2} and imaginary chemical potential $\mu_i$
\cite{fp1,fp2}.

\section{Systematics of reweighting and Taylor expansion}

\begin{figure}[t]
\begin{center}
\includegraphics*[width=0.45\textwidth]{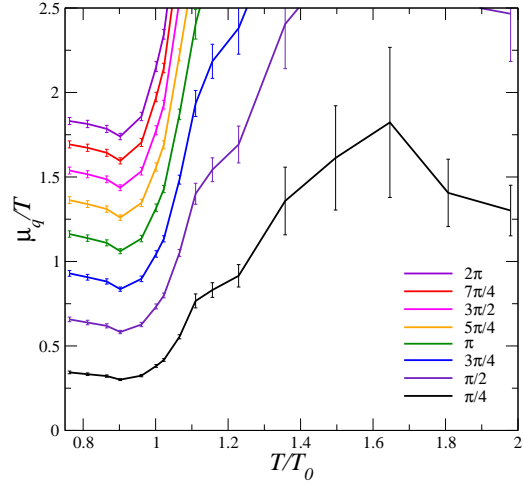}
\caption[]{Contours of $\sigma(\theta)$ for a fixed volume $16^3$ \cite{bisw3}.}
\label{phase}
\end{center}
\end{figure}

A few years into their existence, there are now several investigations of the systematics of the reweighting and Taylor expansion approaches. Recalling the double reweighting formula
\be
\langle O\rangle_{(\beta,\mu)}=
{{\langle O\; e^{{n_{\rm f}\over4}\Delta\ln {\rm det}M}
e^{-\Delta S_g}\rangle_{(\beta_0,0)}}\over
{\langle e^{{n_{\rm f}\over4}\Delta\ln {\rm det}M}
e^{-\Delta S_g} \rangle_{(\beta_0,0)}}}\sim e^{-V\Delta F},
\label{rew}
\ee
one would like to estimate when the exponential suppression of the signal becomes insurmountable.
Splitting the determinant into modulus and phase, $\det M=|\det M| e^{i\theta}$, this should occur when $\langle \cos\theta \rangle\ll1$, or equivalently when the root of the variance of the phase of the determinant grows larger than $\pi/2$,
\be
\sigma(\theta) = \sqrt{\langle \theta^2 \rangle - \langle \theta\rangle^2}= 
\sqrt{\langle \theta^2 \rangle}>\pi/2.
\label{var}
\ee 
In order to quantify this, the Bielefeld-Swansea collaboration evaluated the phase by means of its Taylor expansion \cite{bisw3},
\be
\theta^{(n)}=
{{n_{\rm f}\over4}}\mbox{Im}\sum_{j=1}^n{\mu^{2j-1}\over{(2j-1)!}}
{{\partial^{2j-1}\ln\mbox{det}M}\over{\partial\mu^{2j-1}}}.
\ee
Contours of values for the variance are shown in Fig.~\ref{phase}. According to the criterion Eq.~(\ref{var}), the line corresponding to $\sigma=\pi/2$ can be viewed as the boundary for the reliability of reweighting. The region to its lower right is safe while one would not trust results obtained
up and left from it. This means the deconfined phase of QCD is rather accessible as expected, while in the transition region one finds the constraint $\mu/T_0\lsim 1$, in accord with the constraint on other methods.  
The figure shows contours for one given volume. Reweighting gets exponentially harder with volume, and thus the contour lines move rapidly to the lower right as the volume is increased. 

In another paper \cite{shinji} Ejiri discusses the difficulties of a combined application of reweighting and an analysis of Lee-Yang zeros (LYZ) \cite{lyz}.
The latter exploits the fact that on a finite volume there are no singularities in the pressure, and hence no zeroes of the partition function for real couplings $\beta$. However, there are zeroes for comlex couplings, whose real part indicates the location of an analytic crossover, i.e.~the pseudo-critical temperature $T_0$. In the infinite volume limit, these zeroes move to the real axis if there are true
phase transitions, while they stay at complex values for crossovers.
A LYZ analysis for reweighted finite $\mu$ then amounts to numerically searching for zeros in the expression 
\be
Z_{norm}(\beta_{\rm Re}, \beta_{\rm Im}, \mu)
= \left| 
\left\langle e^{6i\beta_{\rm Im} N_{\rm site} \Delta P} 
e^{i \theta}  
\left| e^{(N_{\rm f}/4) (\ln \det M(\mu) - \ln \det M(0))} \right| 
\right\rangle_{(\beta_{\rm Re}, 0, 0)} \right|,
\label{shin}
\ee
where one additonally reweights into the complex coupling plane.
Ejiri argues in \cite{shinji} that this combined procedure does not have an infinite volume limit.
Taking $V\rightarrow \infty$ at finite statistics, the above expression for the partition function will always go to zero because of the sign problem,
and hence always signal a phase transition, even where there is a crossover. 
This point of principle is not surprising.  Indeed the same mechanism precludes a numerical infinite volume limit of {\it any} observable computed via reweighting. However, the question for practical simulations is whether for a given volume enough statistics can be gathered to beat the sign problem, and whether the volume is large enough to reproduce infinite volume physics with sufficient accuracy. Eq.~(\ref{shin}) illustrates the difficulty of this procedure: the LYZ get masked
by the noise from the reweighting factor, and one has to guard against mistaking a disappearing signal
for a Lee-Yang zero. The problem boils down to being able to give reliable errors for the reweighting procedure, which are needed for a qualified  judgement on whether statistics is sufficient or not. 

\begin{figure}[t]
\includegraphics*[width=0.45\textwidth]{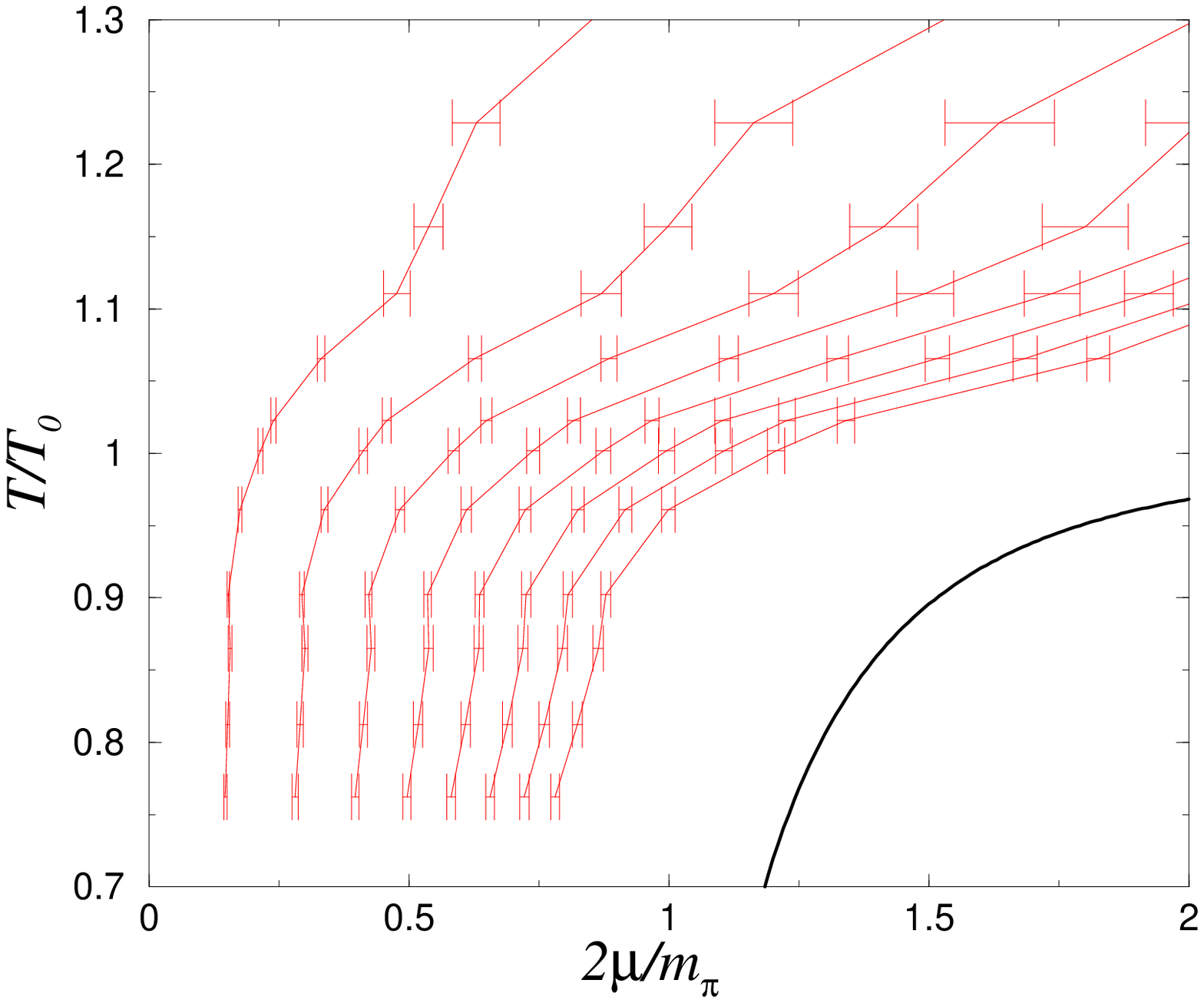}\hspace*{1cm}
\includegraphics*[width=0.45\textwidth]{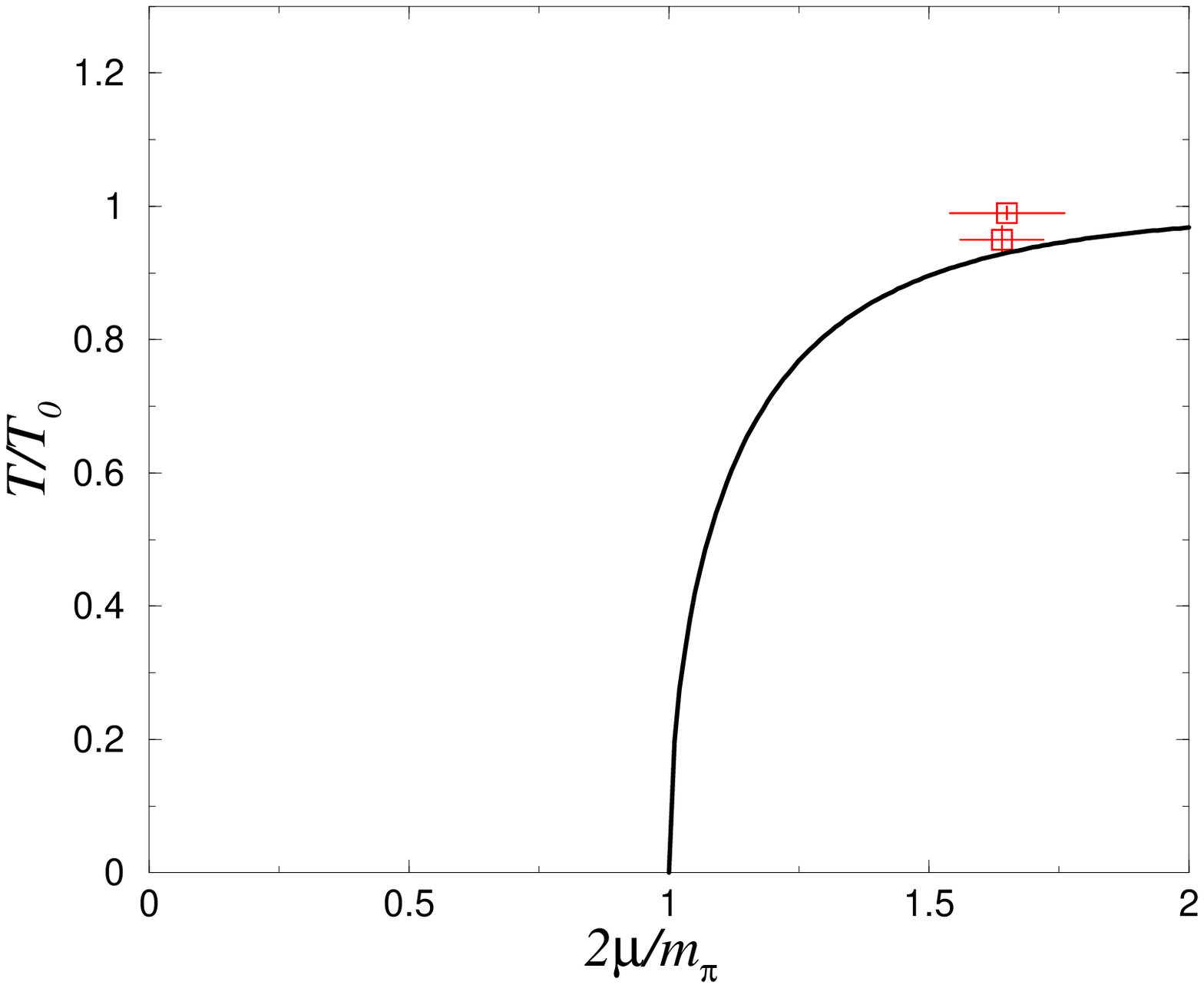}
\caption[]{Left:  rescaled version of the contours of $\sigma(\theta)$, Fig.~\ref{phase}. The values of the contours increase towards the lower right and approach the transition line to the pion condensate at finite isospin density. Right: the critical endpoints for two sets of quark masses by the reweighting method \cite{fk1,fk2}. From \cite{kim}.}
\label{spli}
\end{figure}

In an interesting qualitative investigation of systematics, Splittorff makes use of the finite density formulation \cite{kim}. He suggests to turn the reweighting argument for approximate equality of finite isospin and baryon density around, in order to determine the limit of applicability for reweighting. 
For this purpose a matrix model prediction \cite{mat}
for the transition line to the pion liquid is combined 
with the contour lines for the variance of the phase of the determinant, Fig.~\ref{phase}, as shown in Fig~\ref{spli} (left).
The value of $\sigma(\theta)$ rises towards the lower right, and one observes that contours become denser with larger values, approaching the transition line to the pion liquid. This is to be expected on the grounds that a non-trivial phase of the determinant wipes out the pion condensate, which is not present at finite baryon density. Hence Splittorff suggests to interpret the 
transition line to the pion liquid in the theory at finite isospin density as a ``cut-off'' for the applicability of reweighting: continued reweighting to the right of that separating line would mean one has a serious overlap problem, since the presence of the phase makes a physical difference there. 
That this is a matter worth exploring in more detail is shown in Fig.~\ref{spli} (right), which displays the critical endpoints from reweighting for two different quark mass sets  \cite{fk1,fk2}, and both fall in the neighbourhood of this boundary.

A similar argument may be applied to results from the Taylor expansion. For larger volumes, the sign problem becomes more severe. In the Taylor expansion, whose coefficients are evaluated at $\mu=0$, this shows up in two ways. Firstly the need for more terms to describe the sharpening divergence in its build-up. Secondly, the need for ever more precise cancellations between different terms in order to combine to the correct volume scaling behaviour of the sum. But the severity of the sign problem also puts a limit on $\mu/T$ for fixed volume, as we have seen already. Checking in Fig.~\ref{susc} at which values of $\mu/T_0$ the 
sixth order contribution to the susceptibility becomes important for a given temperature, Splittorff concludes that the 4th order expansion only works to the left of the leftmost contour line in Fig.~\ref{spli} (left). 
Approaches based on imaginary $\mu$ never face the sign problem. However, as dicussed in the next section, analytic continuation to real $\mu$ necessitates a Taylor expansion too, and one would expect a similar limitation. 
Of course, these estimates are not yet quantitative, as the finite isospin transition line is determined from a model and not known with any accuracy, but they point out interesting directions to pursue.

\section{QCD at imaginary $\mu$}

Since the QCD fermion determinant with imaginary $\mu=i\mu_i$ is real positive, it can be simulated just as for $\mu=0$. 
It is then natural to ask whether such simulations can be exploited to learn something about physics at real $\mu$. The strategy to get back to real $\mu$ is to fit the Monte Carlo results, wich are free of approximations, to a Taylor series in $\mu/T$. In case of apparent convergence it is then easy to analytically continue the power series to real $\mu$. 
This idea was first used for observables like the chiral condensate and screening masses in the deconfined phase \cite{lom,hlp}. It was then shown to be applicable to the phase transition itself \cite{fp1}, which has recently been exploited in a growing number of works \cite{fp2}-\cite{cl}.

\begin{figure}[t]
\begin{center}
\includegraphics*[width=0.45\textwidth]{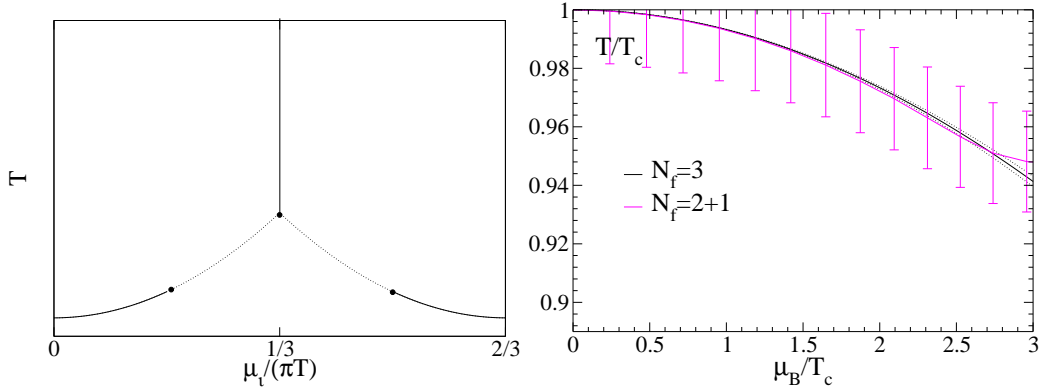}
\includegraphics*[width=0.45\textwidth]{tc_nf_fk.eps}
\caption[]{Left: Schematic phase diagram for QCD at imaginary chemical potential. Right: $N_f=3$ results after continuation \cite{fp2} and compared to $N_f=2+1$ \cite{fk1}. The light quark masses are the same.}
\label{ischem}
\end{center}
\end{figure}

The partition function,
\be
Z(V,\mu,T)=\Tr \left(\ex^{-(\hat{H}-\mu \hat{Q})/T}\right),
\ee
is periodic in the imaginary direction, and the period can be shown to be $2\pi/N_c$ for $N_c$ colours \cite{rw}. Hence, in addition to being even in $\mu$, the QCD partition function has the additional exact symmetry $Z(\mu_r/T,\mu_i/T)=Z(\mu_r/T,\mu_i/T+2\pi/3)$. 
Because of the fermionic boundary conditions in the Euclidean time direction, this symmetry implies that a shift in $\mu_i$ by certain critical values is equivalent to a transformation by the $Z(3)$ centre of the gauge group. Thus, there are $Z(3)$ transitions between neighbouring centre sectors for all $(\mu_i/T)_c=\frac{2\pi}{3} \left(n+\frac{1}{2}\right), n=0,\pm1,\pm2,...$. It has been numerically verified that these transitions
are first order for high temperatures and a smooth crossover for low temperatures \cite{fp1,el1}.
As a consequence, the schematic $(T,\mu_i)$ phase diagram looks as in Fig.~\ref{ischem}. The vertical line coming from the top denotes the $Z(3)$ transition, while the deconfinement transition line now bends upwards as a function of $\mu_i$. The order of the transition and the existence of an endpoint  depends again on the number of flavours and the quark masses. Because of the symmetry of the partition function this picture is then periodically repeated for larger values of $\mu_i$.

The idea then is to simulate with imaginary chemical potential, and fit the full simulation results by
a power series of order $N$,
\be
 \langle O \rangle = \sum_n^N c_n\left(\frac{\mu_i}{T}\right)^{2n}.
 \ee 
Since the Monte Carlo results contain no approximation or truncation, convergence can be inspected
by the quality of the fits to the data. In the case of satisfactory convergence analytic continuation 
$ \mu_i \longrightarrow i\mu_i$ is a trivial matter. It was shown that this strategy can be extended to the pseudo-critical line itself, which on finite volumes is a smooth function with an even Taylor expansion \cite{fp1}. Detailed comparisons give quantitative agreement for $T_0(m,\mu)$ computed from imaginary $\mu$ and other methods \cite{fp2,el1}, cf.~Fig.~\ref{ischem}.
There are also first results for $N_f=4$ with Wilson fermions \cite{cl}.

This procedure may be expected to converge as long as the value of $\mu_i$ does not exceed the critical value of the first $Z(3)$ transition, $|\mu|/T\leq\pi/3$, or $\mu_B\lsim 550{\rm MeV}$ in physical units. This constraint is due to two limitations. Firstly, in the infinite volume limit the location of the $Z(3)$ transitions would bound the radius of convergence of the Taylor series. And secondly even on finite volumes, where there are no non-analyticities, no new information is obtained by going to larger $\mu_i$ because of the periodicity. 
Nevertheless, interesting arguments are being made that one might well extend the continued results along the real $\mu$-axis beyond this radius with the help of, e.g.,
Pad\'e approximants \cite{mar}.

Working at imaginary $\mu$ has a couple of technical advantages. It is computationally simple and much cheaper than reweighting or computing coefficients of the Taylor expansion. Moreover, both parameters $\beta,\mu$ are varied and thus one obtains information from statistically independent ensembles. It also offers some control on the systematics by allowing a judgement on the convergence of the fits.
Furthermore, it is a good testing ground for effective QCD models: analytic results can always be continued to imaginary $\mu$ and be compared with the numerics there, as demonstrated for several examples in \cite{el2}.
The main limitation presently is the radius of convergence in the large volume limit, $\mu/T\sim 1$. 

\subsection{A generalised imaginary $\mu$ approach}

\begin{figure}[t]
\begin{center}
{\rotatebox{270}{\scalebox{0.25}{\includegraphics{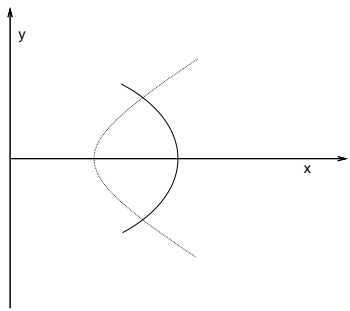}}}}
\caption[]{Schematic phase diagram for QCD at imaginary chemical potential.}
\label{aschem}
\end{center}
\end{figure}

The method of simulating imaginary $\mu$ and analytically continuing can be generalised in an interesting way, as suggested by Azcoiti et al.~\cite{az1}. 
The idea is to rewrite the standard expression for the staggered fermion action at finite density \cite{hk} 
by replacing the chemical potential with
two new parameters $x,y$,
\ba
&& \frac{1}{2}\sum_{n} \bar\psi_n \eta_0 (n)
\left(  e^{\mu a}   U_{n,0}\psi_{n+0}
-  e^{-\mu a}  U^\dagger_{n-0,0}\psi_{n-0}\right)\nn\\
 \rightarrow & &
 x \frac{1}{2} \sum_{n}\bar\psi_n \eta_0 (n)\left( U_{n,0}
\psi_{n+0} - U^\dagger_{n-0,0}\psi_{n-0}\right) 
+  y  \frac{1}{2} \sum_{n} \bar\psi_n \eta_0 (n)\left(  U_{n,0}
\psi_{n+0} + U^\dagger_{n-0,0}\psi_{n-0}\right),
\ea
where $x  =\cosh(a\mu),  y  =\sinh(a\mu)$. This means the action has been enlarged by an extra  parameter. The ordinary finite density action is recovered by the constraint $x^2-y^2=1$. 
Thus, if the solid line in Fig.~\ref{aschem} denotes a phase transition line in the $x,y$ plane of the
enlarged theory, its intersections with the dotted line representing the constraint correspond to physical transition points. The enlarged theory still has the sign problem, but one can simulate at imaginary $y=i\bar{y}$.  The potential of this method to improve over the simple imaginary $\mu$ approach is that
there are now different parameter sets $x,\bar{y}$ to be simulated, so one might hope to be able to extrapolate in a controlled way to reach larger values of $\mu/T$ and thus probe the phase diagram at lower temperatures.

\begin{figure}[t]
\begin{center}
{\rotatebox{90}{\scalebox{0.55}{\includegraphics{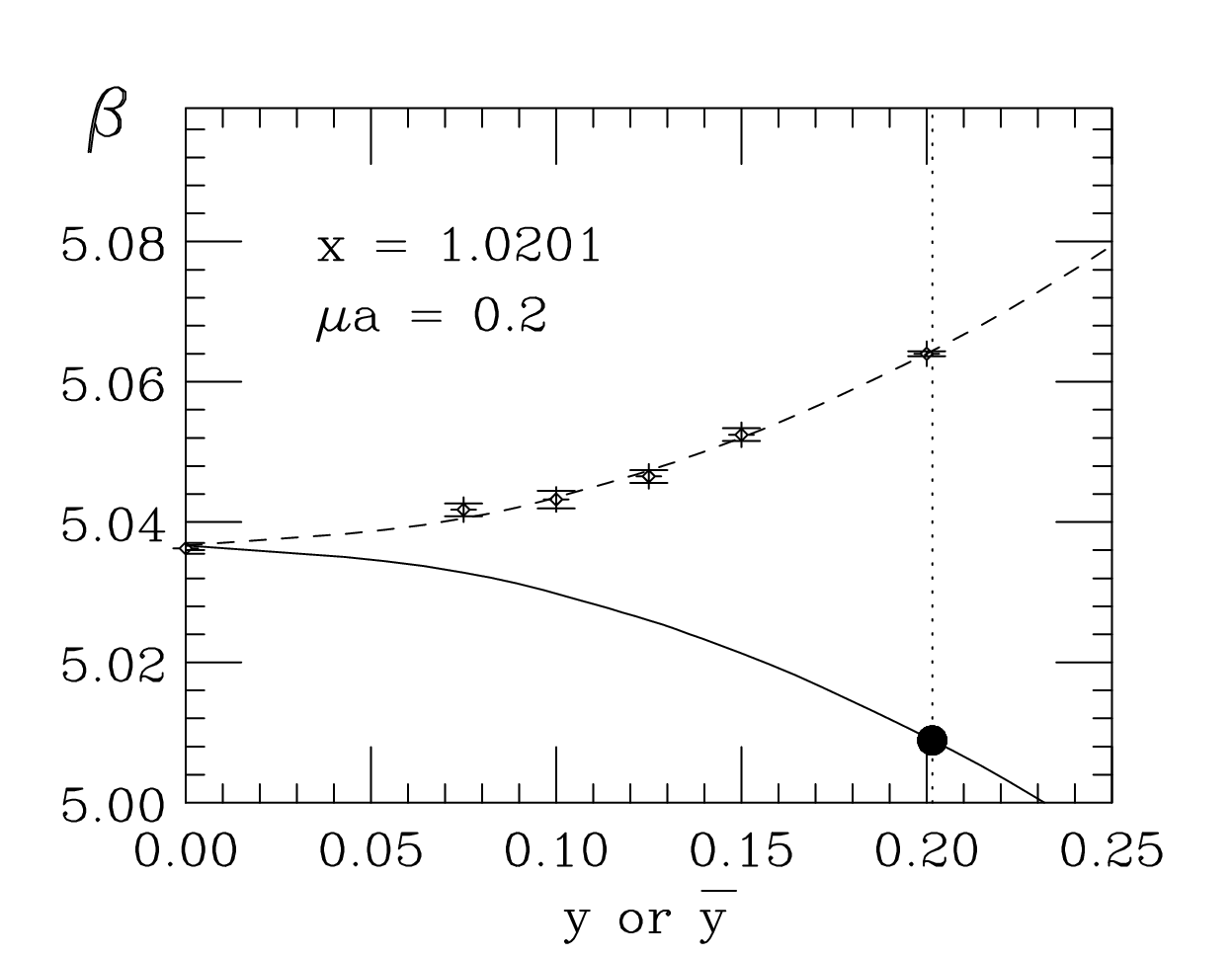}}}}\hspace*{0.5cm}
{\rotatebox{90}{\scalebox{0.55}{\includegraphics{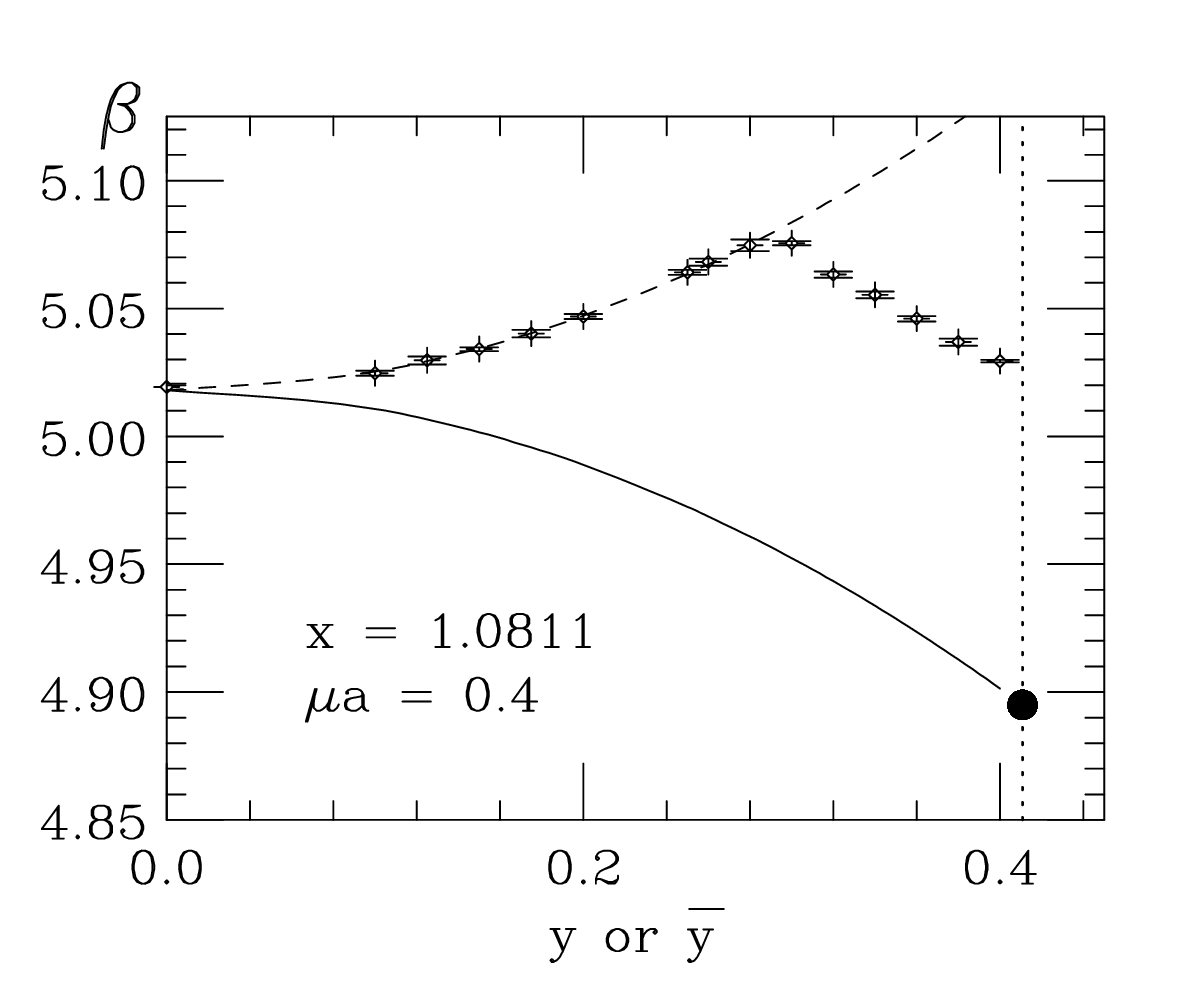}}}}
\caption[]{\label{azres} The critical coupling in the $N_f=4$ theory as a function of imaginary $y$. The leading order quadratic fits are continued to real $y$, the intersection with the vertical line gives the physical value \cite{az2}.}
\end{center}
\end{figure}

Numerical results from this method for the four-flavour theory have been presented in \cite{az2}. Simulations were performed on $8^3\times 4$ lattices with standard staggered fermions and the R-algorithm. Fig.~\ref{azres} shows results for the pseudo-critical coupling obtained at imaginary $y$,
fitted to the form $\beta_c(\bar{y})=\beta_0+\beta_1\bar{y}^2$ and continued to real $y$. The vertical dotted line gives the intersection with physical QCD. Clearly, the method works as well as 
the ordinary imaginary $\mu$ approach. However, its full potential of simulating different sets of $x,y$ for a fixed $\mu/T$ has not been probed yet. Fig.~\ref{azres} (right) shows an example of a result that was extrapolated beyond the first $Z(3)$-transition, which shows up by the kink in the data. This is done on the grounds that no such kink is present in the real direction. Nevertheless, even for imaginary $y$, beyond the kink the curve is not constrained by any additional data points, and hence convergence of the extrapolation is not guaranteed.

New and as yet unpublished results for the two-flavour theory with indications of a critical point are  presented in these proceedings \cite{lal}.
An interesting open question is whether the method can indeed be used to obtain more control over analytic continuation. Another promising idea by the authors is to use their action for reweighting. The two parameters $x,y$ might be tuned such as to shorten the reweighting distance 
to the physical point of interest.  

\subsection{Imaginary $\mu$ and Fourier transformation: QCD at fixed baryon number}

Last year promising attempts of an alternative use of imaginary $\mu$ have been made
\cite{fs,al}. This approach makes use of the relation between the grand canonical partition function at imaginary chemical potential and the canonical partition function at fixed baryon number via Fourier transformation \cite{rw,ht}. After earlier
attempts on a Hubbard model \cite{awk}, there are now promising new results on QCD presented in these proceedings \cite{fs1,al1}. The difficulty in this case is to make baryon number large enough so as to reproduce the finite chemical potential calculations in the thermodynamic limit.

Baryon number is fixed by inserting
$\delta(3 B - \int d^3x ~ \bar\psi\gamma_0 \psi)$ into the path integral. The result is the canonical
partition function, which is related to the grand canonical partition function at imaginary $\mu$ via a Fourier transform,
\be
Z_{C}(B) = \frac{1}{2\pi}
\int_{-\pi}^{\pi} d \left( \frac{ \mu_i }{ T }\right) \; e^{-i 3 B \frac{ \mu_i
}{ T }} Z_{ GC  }(\mu = i   \mu_i).
\ee
The idea followed in \cite{fs} is to sample $Z_{GC}(\mu = i \mu_{MC})$ by Monte Carlo, and then compute
\be
 \frac{Z_{C}(B)}{Z_{GC}(i \mu_{MC})} = 
\left \langle  \frac{1}{\det(i \mu_{MC})}  \int d\mu_i
\exp\left(i 3 B \frac{\mu_i}{T}\right) \det(i \mu_i) \right \rangle,
\ee
i.e.~the fermion determinant gets Fourier transformed, and so it has to be calculated exactly.
This is costly, but the benefit is that now no analytic continuation or Taylor expansion is needed.
The sign problem of course resurfaces here as well, making $Z_C(B)$ noisy. But the strength of this effect is governed by baryon number, and not by volume directly. In the thermodynamic limit one would have to send baryon number to infinity in order to have fixed baryon number density.
Thus, the larger the volume, the smaller the accessible baryon number density for a simulation.
Nevertheless, fixing a small baryon number makes sense in order to study e.g.~nuclear few body systems, and as the following results show, with sufficient computer power one can reach reasonably large baryon numbers to make contact with the grand canonical formulation.

\begin{figure}[t]
{\rotatebox{-90}{\scalebox{0.3}{\includegraphics{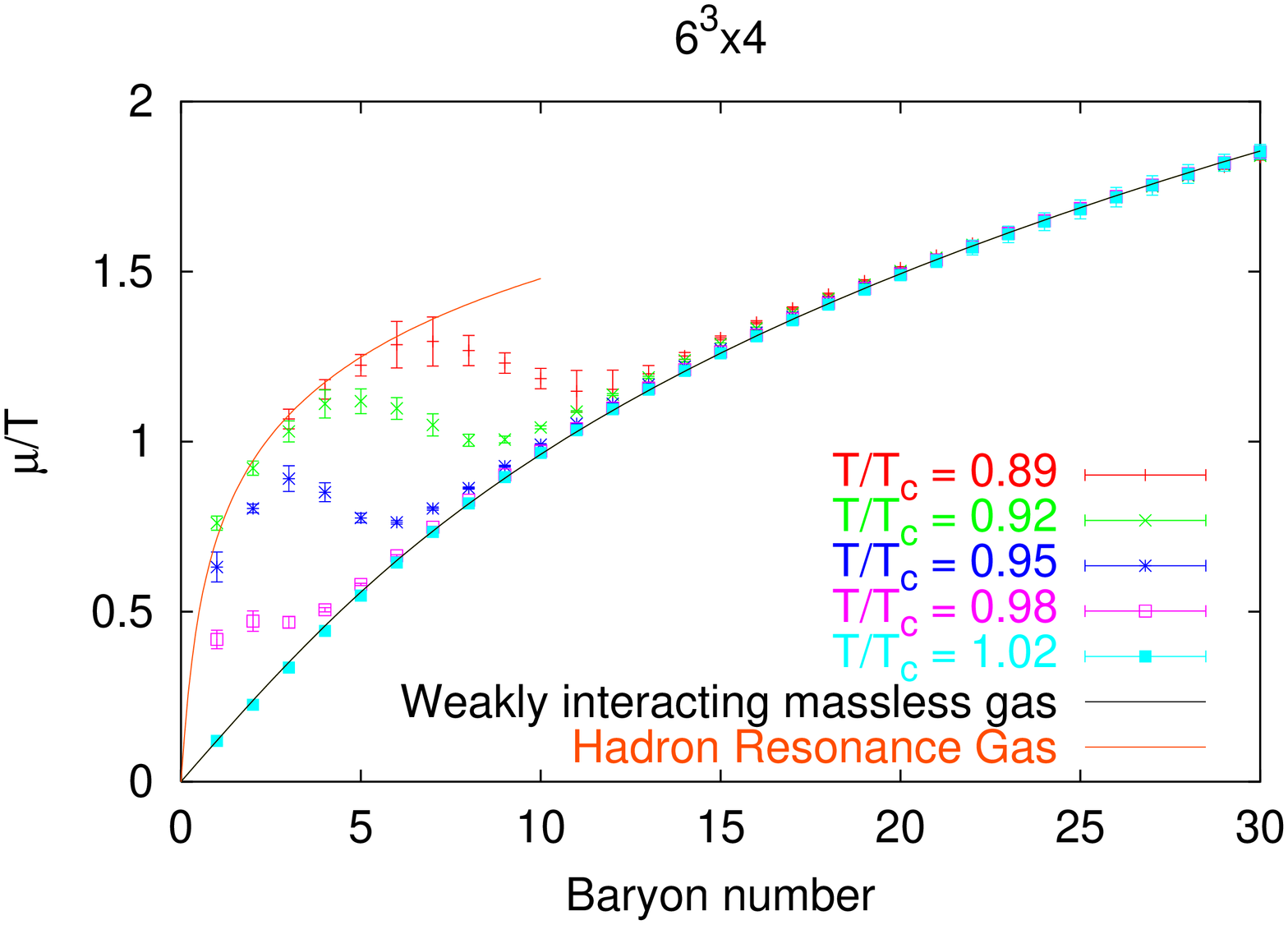}}}} 
{\rotatebox{-90}{\scalebox{0.3}{\includegraphics{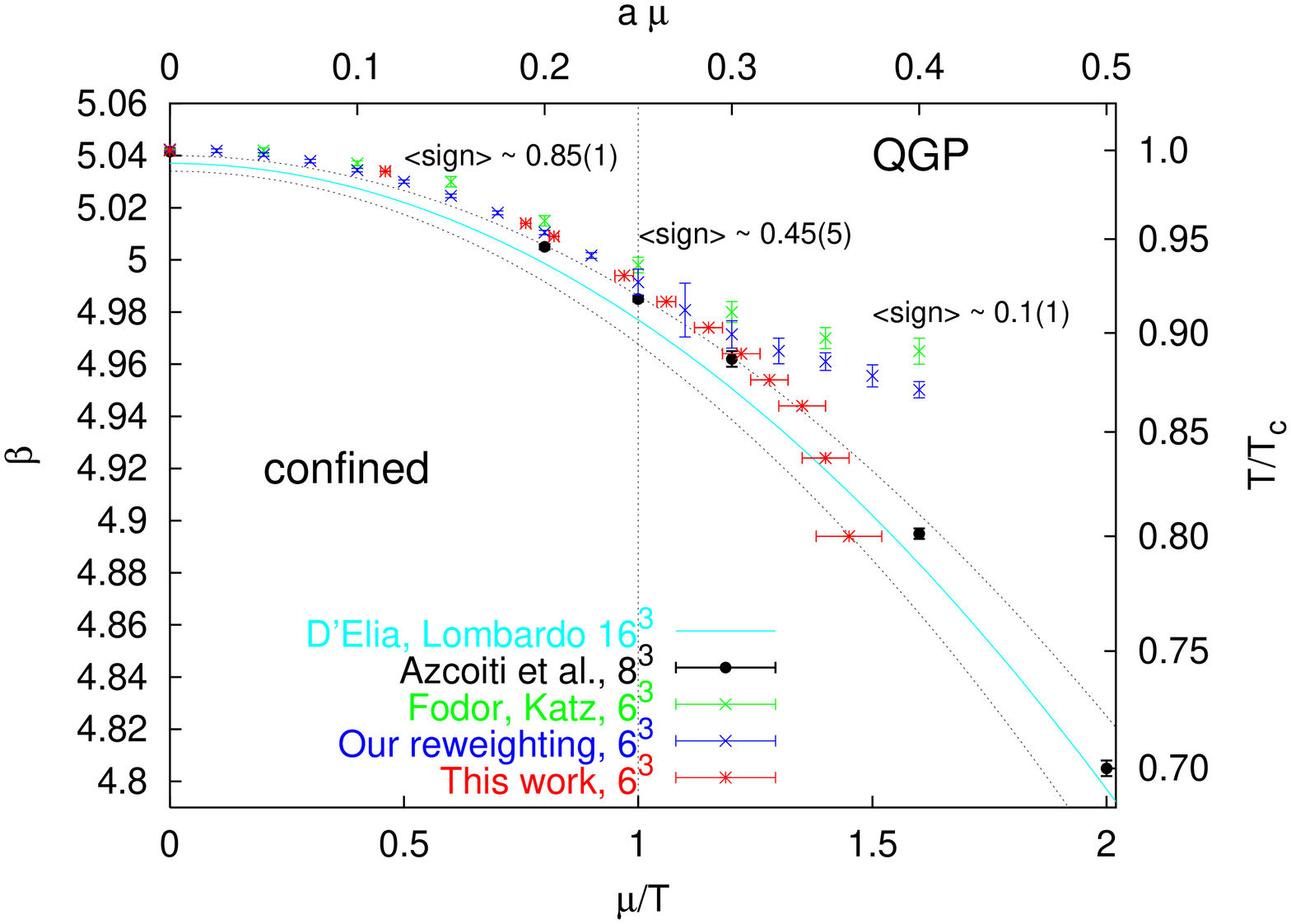}}}} 
\caption[]{Left: Numerical Maxwell construction to relate the canonical and grand canonical ensembles. Right: Comparison of  different methods. Agreement is quantitative for $\mu/T\lsim 1$. From \cite{fs1}.}
\label{can}
\end{figure}

Numerical results obtained by de Forcrand and Kratochvila are shown in Fig.~\ref{can}. They were obtained on a $6^3\times4$ lattice with $N_f=4$ standard staggered fermions and hybrid Monte Carlo simulations, the quark mass was $m/T_0\approx 0.2$. The left panel shows the conversion from 
baryon number to chemical potential. This is achieved by evaluating the free energy $F(B)$ as function of different fixed baryon numbers, and computing the grand canonical partition function by Laplace transformation. The integral can be evaluated by means of 
a saddle point expansion,
\be
Z_{GC}(\mu) = \int d\rho \exp\left(-\frac{V}{T}(f(\rho) - \mu  \rho)\right),
\ee
yielding the chemical potential  as a function of baryon number,
\be
\mu  \approx f'(\rho) \approx  \frac{F(B+1)-F(B)}{3}.
\ee

As Fig.~\ref{can} (left) shows, on a $6^3$ lattice it is possible to perform this up to quite respectable baryon numbers. The resulting picture essentially shows the Maxwell construction for changing between canonical and grand canonical ensembles. The S-shaped curves are indicative of a first order phase transition and represent the metastability region. The two envelopes can be ascribed to the hadronic and quark gluon phases and are well fitted by hadron and weakly interacting massless gas models.
Preliminary results from the canonical ensemble obtained with Wilson fermions have been published
in \cite{al} and are presented in these proceedings \cite{al1}.

Fig.~\ref{can} (right) shows the resulting critical line for the four flavour theory in comparison with ones obtained by other methods discussed here. All data are generated with the same action and for the same quark mass and lattice spacing, only the volumes differ between the data sets. For $\mu/T\lsim 1$, the quantitative agreement is impressive. Only the data from \cite{el2} are somewhat off the others,  presumably due to the much larger volume of that data set. Note that for $\mu/T>1.3$ agreement stops and the different 
data sets diverge. This is in accord with the previous statements that all methods
discussed have roughly the same range of applicability, but different systematics. 
The data continued from imaginary $\mu$ in this region are essentially unconstrained and just extrapolated. For the reweighted data points the expectation value of $\cos\theta$ is quoted in the figure for selected points. For $\mu/T>1.2$ it is completely lost in the noise and hence the data points
are not trustworthy. Note also that the data coming from the canonical ensemble in principle do not 
have the restriction $\mu/T\lsim 1$. The data bend down more strongly as  one might expect in this region of the phase diagram. It will be exciting to see whether the curve can be reliably continued beyond $\mu/T\sim 1$. With the density of states method \cite{cs}, Pad\'e approximants \cite{mar} and
the canonical ensemble \cite{fs}, there are at least three attempts at work in this direction.  

\section{The critical end point and its quark mass dependence for three flavours}

\begin{figure}[t]
\vspace*{0.3cm}
{\rotatebox{0}{\scalebox{0.3}{\includegraphics{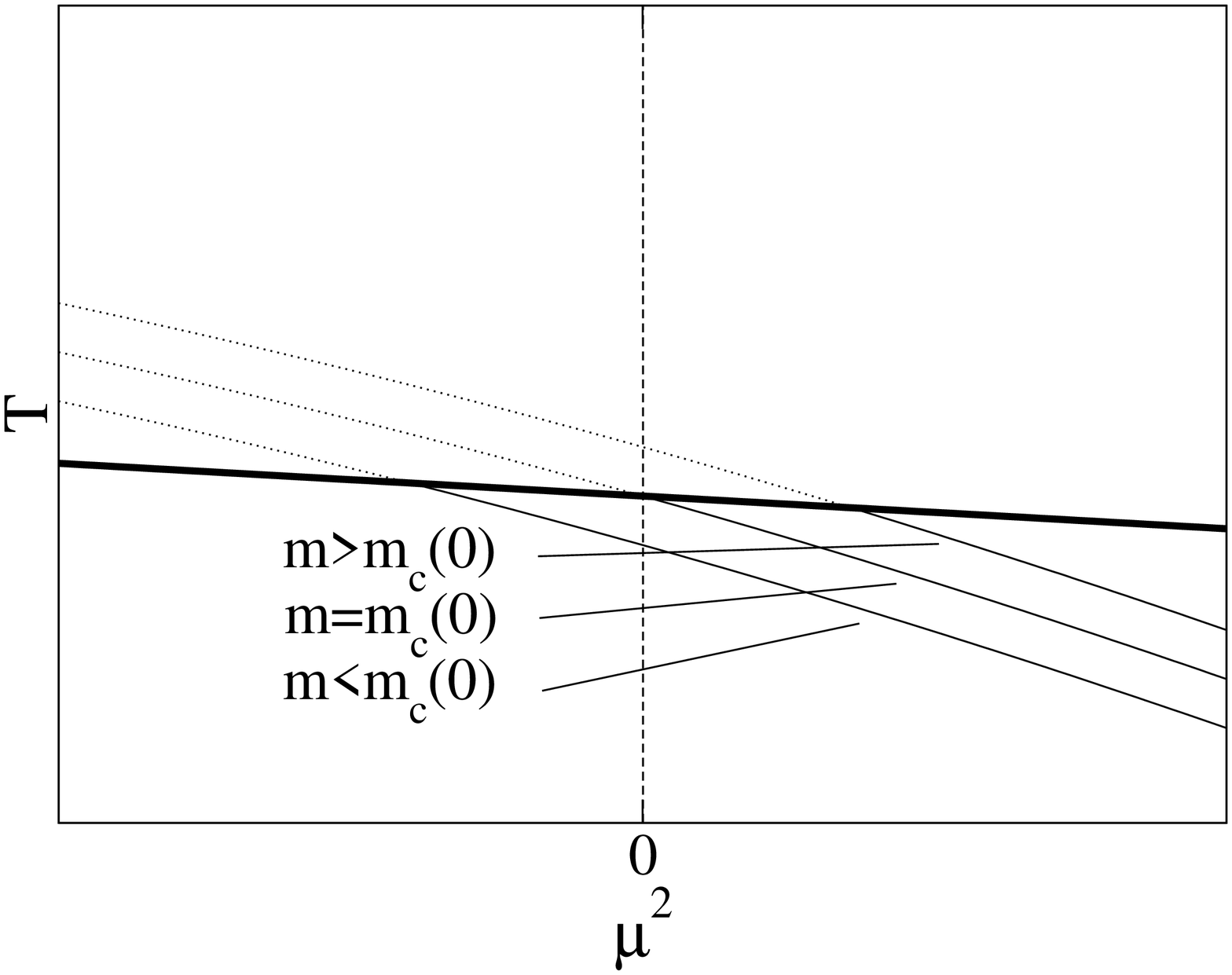}}}} \hspace*{0.5cm}
{\rotatebox{0}{\scalebox{0.3}{\includegraphics{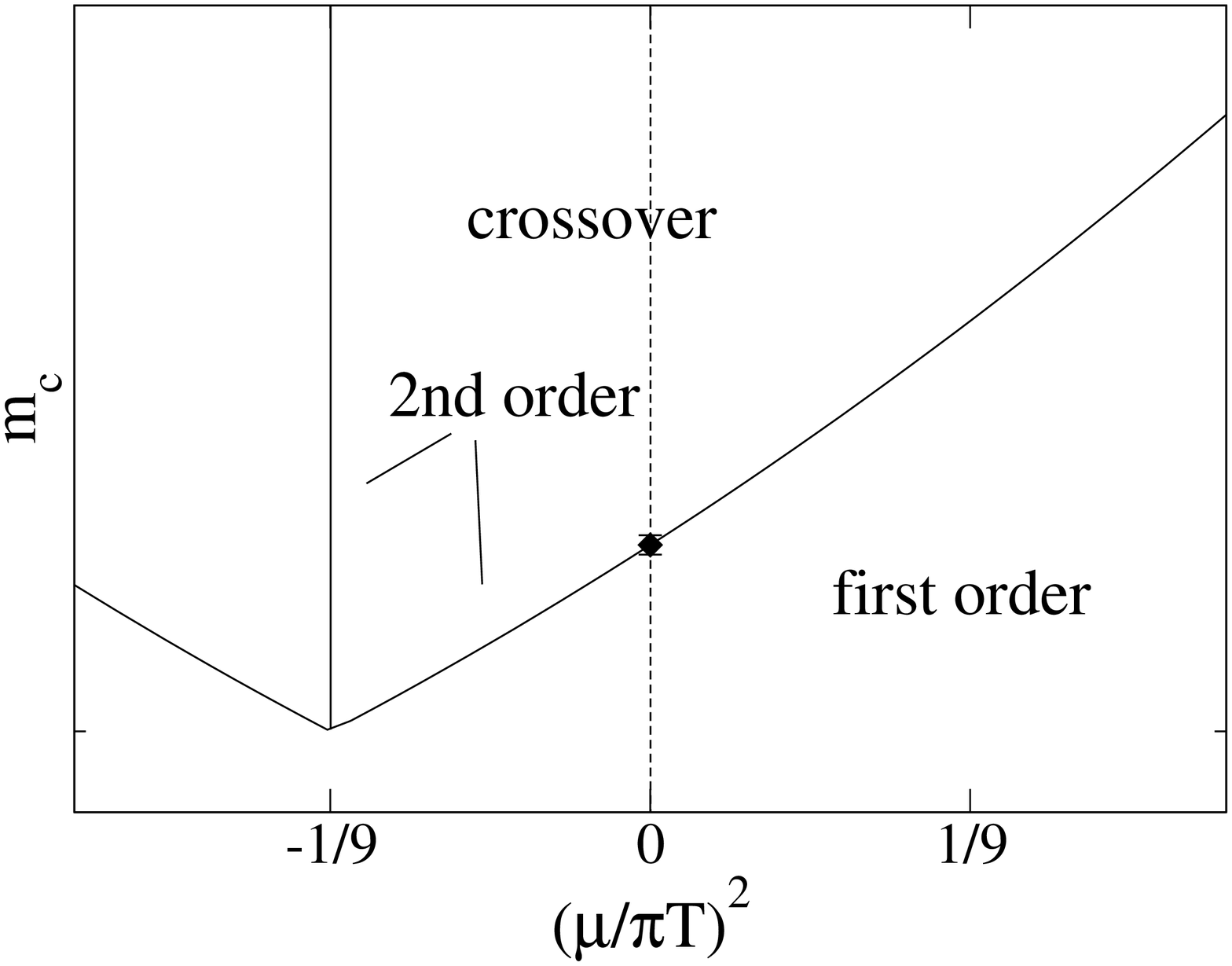}}}} 
\caption[]{Left: Schematic phase diagram in the $(T,\mu^2)$-plane for different quark masses. Solid lines are first order, dotted lines crossover, and the thick line represents the endpoints. Right: The critical quark mass as 
a function of $\mu^2$. The vertical line on the left marks the first $Z(3)$ transition at imatinary $\mu$. From \cite{fp2}. }
\label{tcschem}
\end{figure}

As has become clear by now, a determination of
the order of the phase transition and the 
critical point is much more demanding than the location of the pseudo-critical temperature $T_0(m,\mu)$.
The best starting point for such an enterprise is the $N_f=3$ theory. This is because we know there is a critical point at $\mu=0$. In the phase diagram Fig.~\ref{schem_2+1} (left), the critical quark mass $m_c(\mu=0)$ separating the crossover from the
first order sections along the three flavour diagonal is known to be at a moderately small value accessible to simulations \cite{kls,clm,fp2}.  With the quark mass tuned to this value,
the first order transition line in the $(T,\mu^2)$ phase diagram Fig.\ref{tcschem} (left) reaches all the way to the temperature axis, on which it ends with a critical endpoint. According to the standard scenario discussed in the introduction, if we now increase the quark mass to values $m_c>m_c(0)$, the whole transition line shifts, with the critical endpoint wandering to the right towards a real $\mu_c\neq 0$, thus tracing out a smooth function $\mu_c(m)$. On the other hand, if we change the quark mass to $m<m_c(0)$, the critical endpoint should wander in the imaginary $\mu$-direction to the left, as shown
in Fig.~\ref{tcschem}.

Inverting the function $\mu_c(m)$, we are interested how the critical quark
mass separating first order from crossover changes with $\mu$. This function again has a Taylor expansion in even powers of $\mu$, and one expects
\be
\frac{m_c(\mu)}{m_c(\mu=0)}=1 +  c_1 \left(\frac{\mu}{\pi T}\right)^2+\ldots
\label{c0}
\ee
with coefficients of order one.
In the three-dimensional phase diagram of Fig.~\ref{schem_2+1} (right) this means
we are looking for the curvature of the critical surface in the $N_f=3$ direction at $m_c(0)$. Once this functional dependence is determined, it will return the critical end point $\mu_c$ for a given quark mass.

\subsection{Numerical results for $N_f=3$ from imaginary $\mu$ \label{sec:3f}}

\begin{figure}[t]
\begin{center}
\includegraphics*[width=0.45\textwidth]{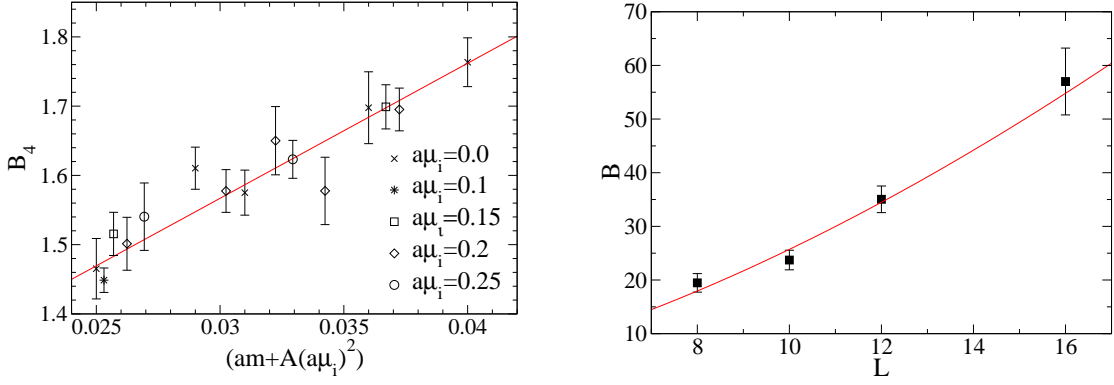}\hspace*{1cm}
\includegraphics*[width=0.45\textwidth]{scale.eps}
\caption[]{Left: The Binder cumulant calculated for various masses and chemical potentials. The value 1.604 corresponds to a critical point. Right: Finite volume scaling $\sim L^{1/\nu}$ of the fit parameter $B$, Eq.~(\ref{bfit}).  $\nu=0.62(3)$ is consistent with the universal Ising exponent $\nu=0.63$ \cite{fp2}.}
\label{b4}
\end{center}
\end{figure}

This program was recently carried out in \cite{fp2}, using $8^3\times4$ lattices with the standard staggered action and the R-algorithm. The observable used to find the critical point was the Binder cumulant, making use of the knowledge that the transition is in the universality class of the 3d Ising model. For a critical point in this universality class the value of $B_4$ is accurately known,
\be
B_4(m_c,\mu_c)=\frac{\langle(\delta\bar{\psi}\psi)^4\rangle}
{\langle(\delta\bar{\psi}\psi)^2\rangle^2}\rightarrow 1.604,\quad V\rightarrow \infty,
\ee
while $B_4\rightarrow 1(3)$ for a first order transition (crossover). Hence in the infinite volume limit $B_4$ is a non-analytic step function. However, on finite volumes it will pass through the Ising value
smoothly, with a slope increasing with volume. Measurements for several values of imaginary $\mu$ and quark masses are shown in Fig.\ref{b4} (left). The data can be fitted to a leading order Taylor expansion in both the quark mass and chemical potential about the known critical point at $m_c(\mu=0)$,
\be
B_4(am,a\mu)=1.604 + B\left(am-am_c(0) + A(a\mu)^2\right) + \ldots
\label{bfit}
\ee
From the fit parameters one can directly extract the desired coefficient $d(am)/d(a\mu)^2$ in lattice units.
For the continuum conversion one needs to take into account that the lattice spacing effectively changes
with $T(\mu)$, and hence $a(\mu)\neq a(0)$. For $c_1$ in Eq.~(\ref{c0}) this yields
\be
c_1=\frac{1}{m_c(0)}\frac{dm_c}{d(\mu/\pi T)^2}=\frac{\pi^2 }{N_t^2(am_c)(0)}\frac{d(am_c)}{d(a\mu)^2}
+\frac{1}{T_0}\frac{dT}{d(\mu/\pi T)^2},
\label{conv}
\ee
Fig.~\ref{b4} (right) shows a finite size scaling analysis of the fit coefficient $B$, and the fitted volume scaling nicely reproduces the exponent predicted by universality even on moderate volumes.
Note that these are extremely difficult calculations, as very long Monte Carlo trajectories of order 80k are required in order to get sufficient tunneling statistics.

In accord with qualitative expectations, $c_1\approx 0.8(4)$ \cite{fp2} is of order one.  However, the large statistical error indicates that the coefficient is also consistent with being close to zero. Independent of the accuracy of the result,  
an important conclusion is that $m_c$ changes very little when 
$\mu$ is switched on, or conversely that $\mu_c$ changes rapidly under small variations of the quark mass.

\subsection{Numerical results for $N_f=3$ at finite isosipin density}

\begin{figure}[t]
\begin{center}
\includegraphics*[width=0.45\textwidth]{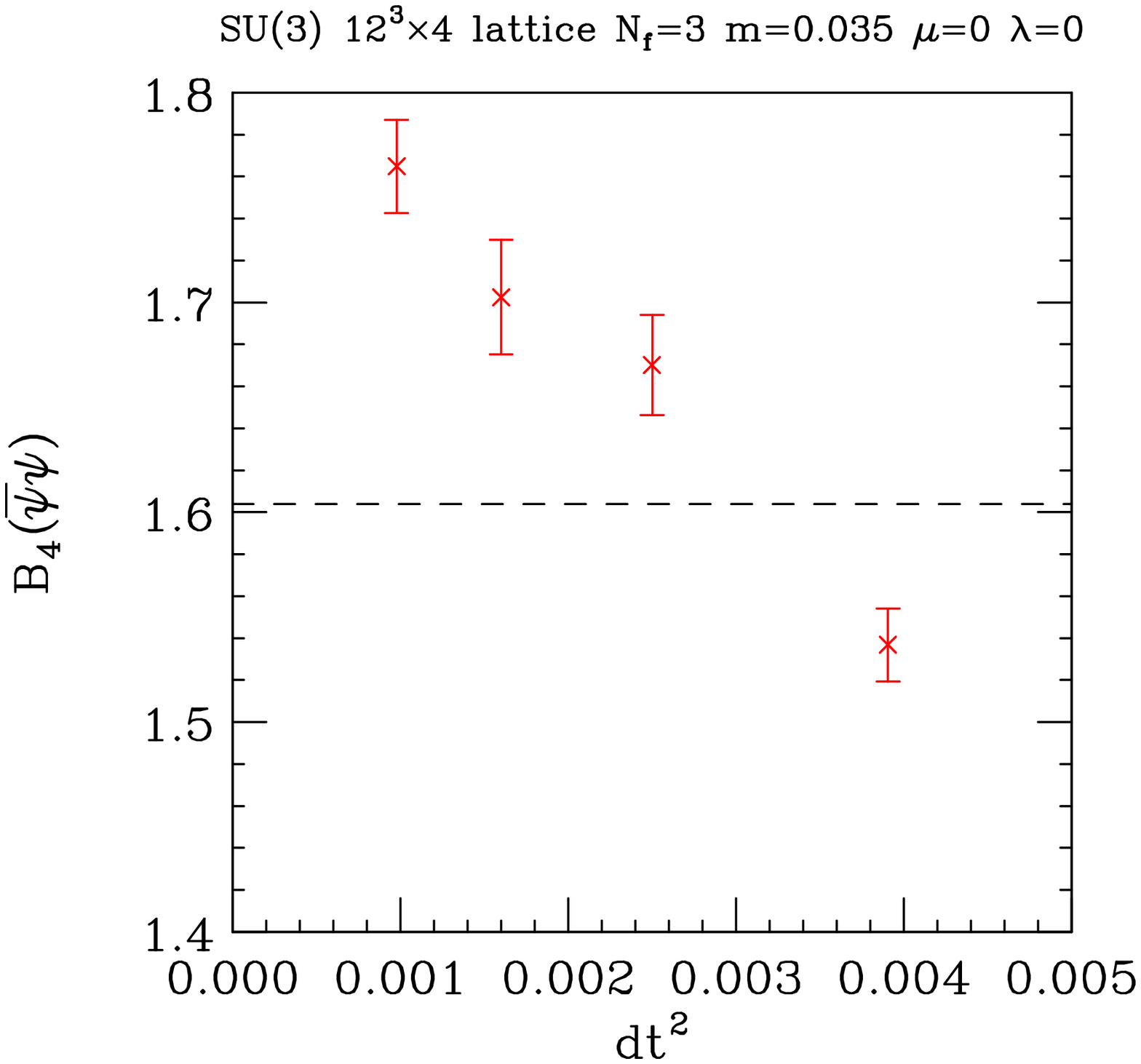}\hspace*{1cm}
\includegraphics*[width=0.45\textwidth]{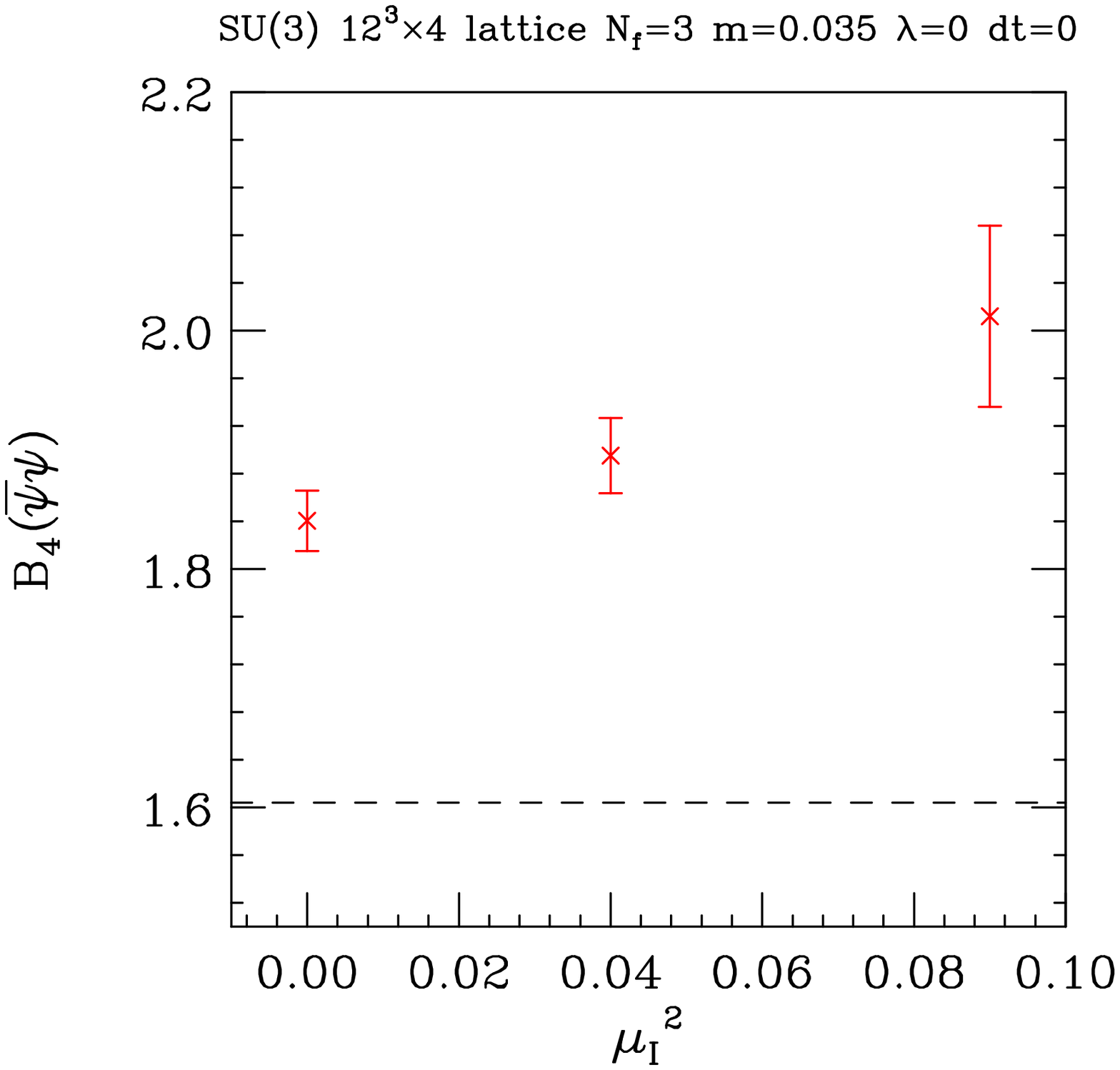}
\caption[]{Left: Stepsize dependence of the Binder cumulant. Right: Binder cumulant extrapolated to 
zero stepsize. Dependence on isospin is very weak \cite{ks2}}
\label{step}
\end{center}
\end{figure}

In a recent article Kogut and Sinclair report on similar investigations in the theory at finite isospin density \cite{ks2}. They work on $L^3\times4$ lattices with $L=8-16$, using the standard staggered action and the R-algorithm. Their quark mass is chosen as $m\gsim m_c(0)$, with the aim to see how the critical point moves as a function of $\mu_I$. As an observable they use the Binder cumulant discussed in the previous section.
Results from their simulations are shown in 
Fig.~\ref{step}. 
The left panel displays an investigation of step-size effects on the Binder cumulant, which are found to be significant in this quark mass regime. Instead of extrapolating to zero step size, the standard usage of  the R-algorithm is to simulate at some reference step size whose error is known to be smaller than typical statistical errors at some reference mass $m$. When going to smaller quark masses, a common practice is to keep the step size to be half the bare quark mass. However, in the low quark mass regime of interest here, this procedure clearly breaks down, with even qualitative changes of the results. The figure shows how at the same quark mass the transition looks clearly first order for large step sizes, but changes to crossover behaviour once extrapolated to zero stepsize, which thus is mandatory.

The right panel shows the results after such extrapolations have been performed. In agreement with the findings from imaginary $\mu$ in the previous section, the Binder cumulant is practically flat and only very weakly depending on $\mu_I$.  All data points are well in the crossover regime. What is surprising is that the weak $\mu_I$-dependence has a tendency for $dB_4/d\mu_I^2\gsim 0$. This would imply that  as $\mu_I$ is switched on the transition moves deeper into the crossover regime, instead of approaching a critical point!

\subsection{$N_f=3$ and $N_f=2+1$ with an exact algorithm \label{sec:line}}

\begin{figure}[t]
\begin{center}
\includegraphics*[width=0.45\textwidth]{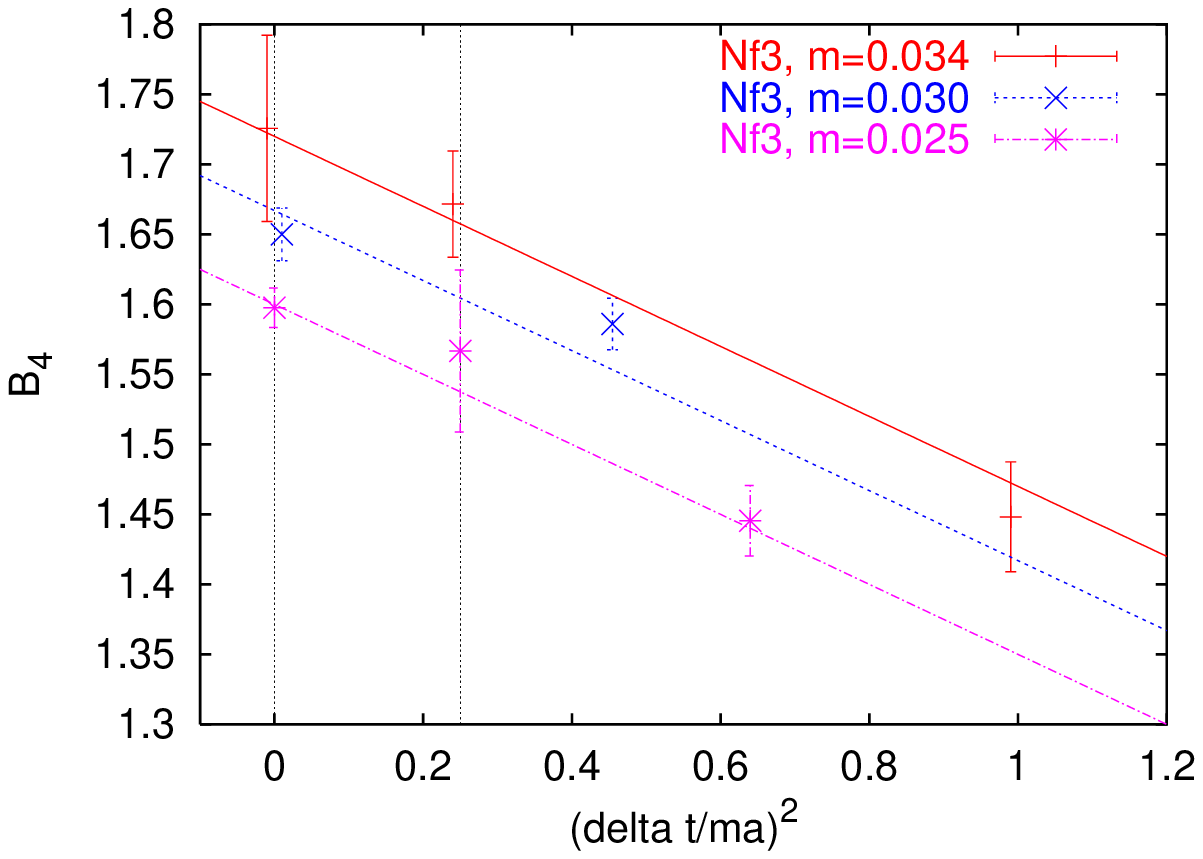}\hspace*{1cm}
\includegraphics*[width=0.45\textwidth]{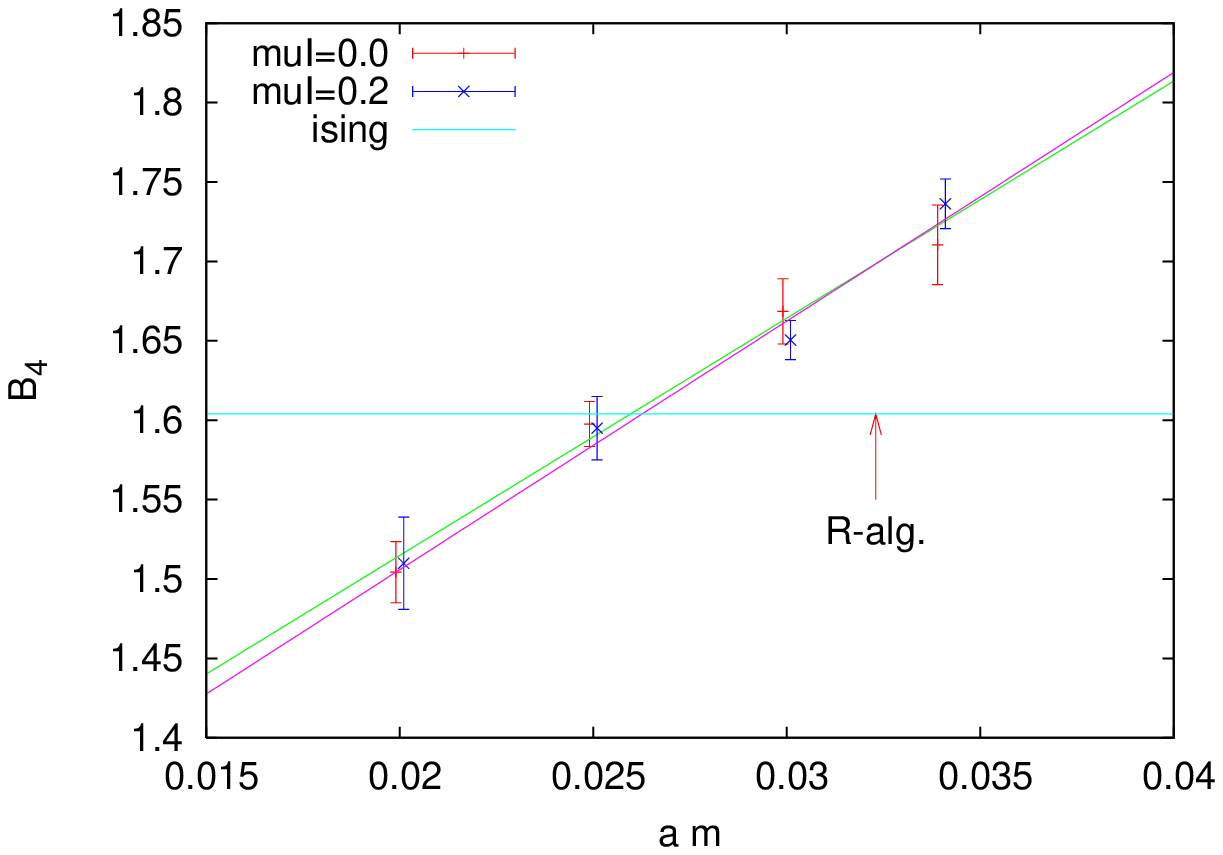}
\caption[]{Left: Comparison of the Binder cumulant computed with the RHMC algorithm (leftmost data) and the zero stepsize extrapolation of the R-algorithm. Right: Determination of $m_c(\mu)$ for $\mu=0,0.2$ with the RHMC algorithm. The arrow marks the result from the R-algorithm.}
\label{rhmc}
\end{center}
\end{figure}

In order to check for stepsize effects and clarify the sign of the $\mu$-dependence of $B_4$,
de Forcrand and I have redone the $N_f=3$ calculation of \cite{fp2} reported in Section  \ref{sec:3f} with 
the exact rational hybrid Monte Carlo (RHMC) algorithm developed by Clark, Kennedy and Sroczynski \cite{rhmc}
(also presented in these proceedings \cite{clark}).
In a first test we compared simulations of the Binder cumulants performed with that algorithm to ones from the R-algorithm extrapolated to zero stepsize. The results are shown in Fig.~\ref{rhmc} (left), and perfect agreement is found.
The right panel then shows a new determination of the critical quark mass $m_c(\mu)$, both for zero density and an imaginary chemical potential, using the RHMC algorithm. Note that the Binder
cumulant now passes its Ising value at a significantly different mass compared to the results in the literature. We find $am(0)\approx 0.026$, which is a shift of about 25\% due to stepsize effects!
On the other hand, switching on a chemical potential has no effect on the Binder cumulant. As a preliminary result we find
$d(am_c)/d(a\mu)^2\approx 0$ within errors, which
is consistent with the findings at finite isospin \cite{ks2} reported in the last section. 

\begin{figure}[t]
\begin{center}
\includegraphics*[width=0.45\textwidth]{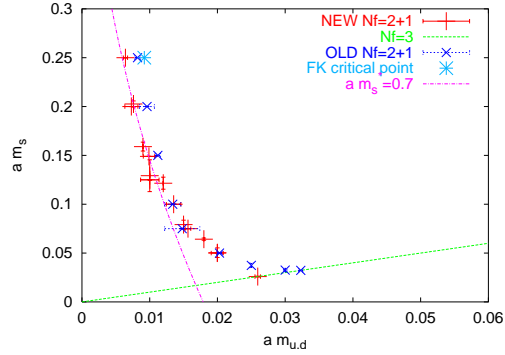}
\caption[]{The critical line separating first order from crossover for $N_f=2+1$. A significant shift is observed when eliminating stepsize effects. The line passes in close vicinity of the physical point (FK).}
\label{crit}
\end{center}
\end{figure}

We have also mapped out the critical line for non-degenerate quark masses, as shown in Fig~\ref{crit},
both with the R-algorithm and the RHMC algorithm. In general there is a significant step size effect.
The picture clearly puts the physical point, where Fodor and Katz performed their simulations, on the crossover side of the line. Note, that the physical point is very close to the critical line.
This is consistent with the requirement of finely tuned quark masses in order to have a critical point at moderate chemical potentials. (The calculation of $c_1$ in this case is still in progress. Taking our R-algorithm result $c_1$ from the three flavour case, our resulting $\mu_c$ would be consistent with theirs within errors). If the chiral limit of the two flavour theory turns out to be O(4), there is a tri-critical point
at some quark mass $m_s^{tric}$ on the $m_s$-axis. Our results can be fitted with the corresponding 
scaling equation and would then predict $m_s^{tric}/T_0\approx 2.8$.

\subsection{A non-standard scenario for the phase diagram}

\begin{figure}[t]
\vspace*{-0.5cm}
\begin{center}
\includegraphics*[width=0.5\textwidth]{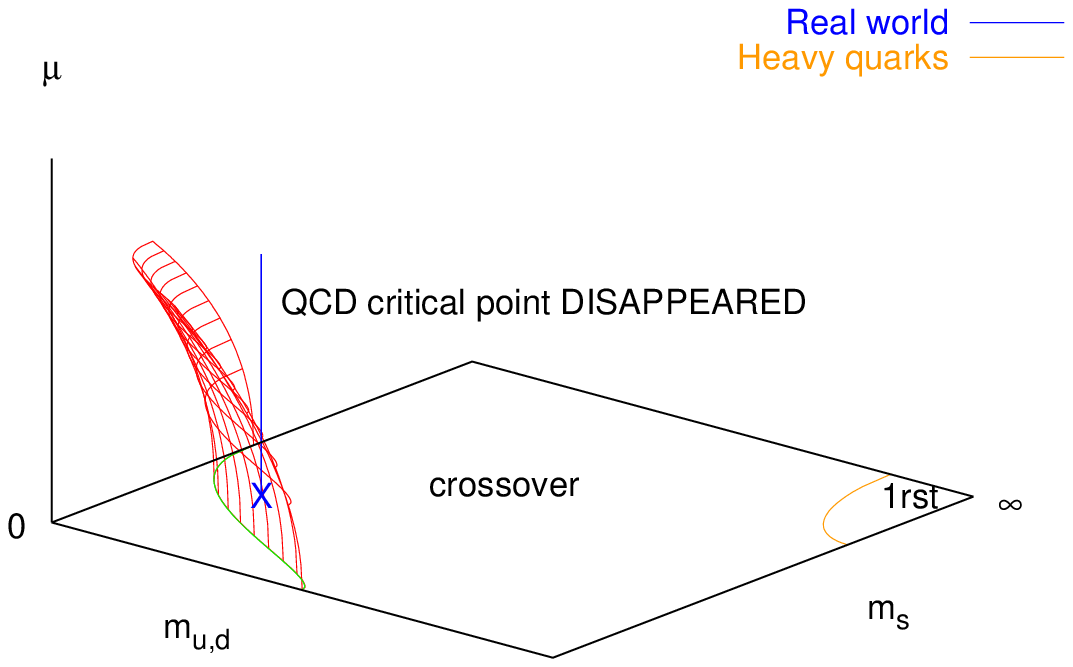}\hspace*{1cm}
\includegraphics*[width=0.4\textwidth]{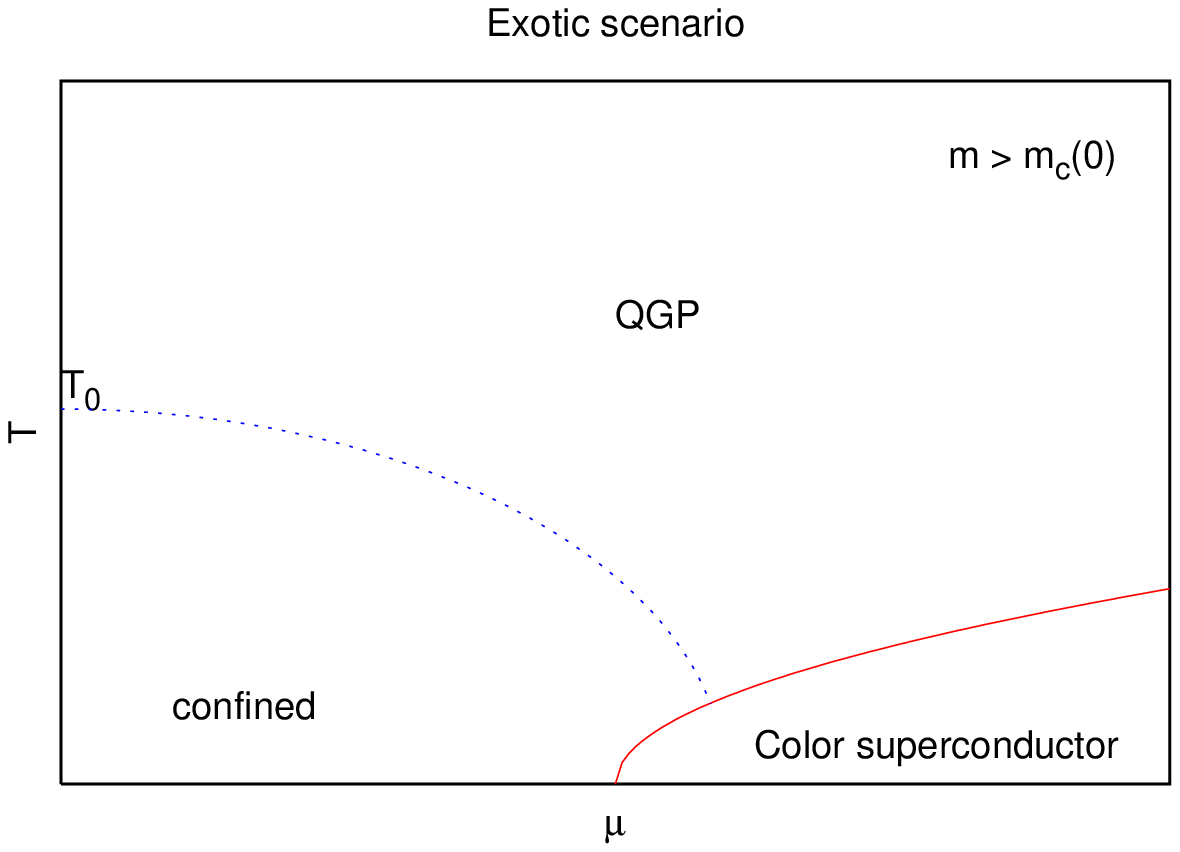}
\caption[]{
For $dm_c(\mu)/d\mu^2 <0$, there is no critical point at all, the dotted line on the right is merely a crossover.}
\label{nons}
\end{center}
\end{figure}

Let us assess the consequences of the step size effects on the critical point. 
After continuum conversion the new result for $m_c(\mu)$, now free of step size errors, is
\be
\frac{m_c(\mu)}{m_c(\mu=0)}=1  -  0.6(2) \left(\frac{\mu}{\pi T}\right)^2.
\ee
Note that the sign of the leading term has changed compared to the previous result! 
The reason is that the first term in Eq.~(\ref{conv}) now is consistent with zero, so the negative second term dominates. 
This means the critical mass gets smaller when a real $\mu$ is switched on, and hence that the critical surface in the phase diagram leans towards smaller quark masses, Fig.~\ref{nons}, i.e.~the 
opposite of the standard scenario Fig.~\ref{schem_2+1} (right). 
In this case, the first order region in a plane of constant $\mu$ is actually shrinking for growing $\mu$. If the physical point is in the crossover region at $\mu=0$, then switching on a chemical potential will not lead to an intersection with the critical surface, and hence there would be no critical point or first order phase transition at all!
Note that this scenario is perfectly consistent with all the universality arguments summarised in Section \ref{sec:qual}. The $(T,\mu)$ phase diagram would then only have the transition line separating the superconducting phase from nuclear matter, as in Fig.~\ref{nons}.

\subsection{Can one expect a critical end point at $\mu_B\lsim 500$ MeV?}

Clearly, as the discussion of the systematics in the last section revealed, this last result is preliminary as well, an important question being in which direction corrections go as the continuum limit is approached.
Nevertheless, based on the results on the curvature of the critical surface, one may obtain a rough estimate for the conditions required to have a critical endpoint in the phenomenologically interesting region $\mu_B\lsim 500$ MeV. Irrespective of whether an eventual continuum result for $m_c(\mu)$ will have positive or negative curvature,  as long as the coefficient $c_1$ in Eq.~(\ref{c0}) is $\sim O(1)$ ( its  natural size; all known coefficients in the pressure \cite{bisw3}, screening masses \cite{hlp} and the pseudo-critical temperature \cite{fp2} are of that order), it implies a very strong quark mass dependence of the value of $\mu_c$. For instance, in order to have $\mu_c\sim 120$ MeV as predicted
by Fodor and Katz \cite{fk2} for $N_f=2+1$, the quark mass has to obey the condition
\be
1<\frac{m}{m_c(\mu=0)}\lsim 1.05,
\ee
i.e.~it has to be fine-tuned to be within 5\% of the critical quark mass. Provided the coefficient $c_1$
does not change drastically in the case of $N_f=2+1$, a similar situation will be encountered there as well. 

At this point it should become clear that we are still far from a quantitative solution of the problem of the critical endpoint. To achieve a resolution better than $5\%$ in the quark masses would even require
to distinguish the up and down quarks. By contrast, we have just discovered a 25\% systematic
error in the critical quark mass due to step size effects, and we have discussed earlier the strong cut-off dependence ($\sim 100\%$) of the critical quark masses in physical units. Thus we should expect formidable shifts in $\mu_c$ on the way to a reliable continuum result.  

\subsection{The heavy quark limit: Potts model}

Other interesting projects are concerned with the upper right corner of Fig.~\ref{schem_2+1}, i.e.~the region towards the quenched limit. 
Simulations of quenched QCD at finite baryon number have been done in \cite{quenb}. 
As the quark mass goes to infinity, quarks can be integrated out and QCD reduces to a gauge theory of Polyakov lines. First simulations of this theory with Wilson valence quarks can be found in \cite{nucu}.
At a second order phase transition, universality allows us to neglect the details of gauge degrees of freedom, so the theory should be in the universality class of the 3d three-state Potts model, which is the 3d Ising model. Hence, studying the three-state Potts model should teach us about the behaviour of QCD in the neighbourhood of the critical line separating the quenched first order region from the crossover region.
For large $\mu$ the sign problem in this theory was actually solved by means of cluster algorithms recently \cite{clust}. 

\begin{figure}[t]
\begin{center}
\includegraphics*[width=0.45\textwidth]{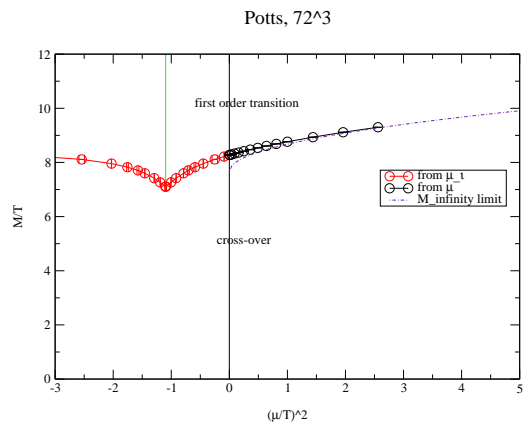}
\caption[]{The critical heavy quark mass separating first order from crossover as a function of $\mu^2$ \cite{skim}.}
\label{potts}
\end{center}
\end{figure}

Here I want to discuss an as yet unpublished result on simulations of the three state Potts model as presented in these proceedings \cite{skim}. For small $\mu/T$, the sign problem of this theory is mild enough so that brute force simulations at real $\mu$ are feasible. In the simulations presented in \cite{skim}, the change of the critical heavy quark mass is determined as a function of real as well as imaginary $\mu$, as shown in Fig.~\ref{potts}. 
Note that $M_c(\mu)$ rises with real chemical potential. i.e.~the first order region in Fig.~\ref{schem_2+1} shrinks as finite baryon density is switched on. This system is thus an example of the non-standard scenario discussed in the previous sections! 
Note also that the qualitative behaviour in going from real to imaginary $\mu$ is exactly as predicted in
the schematic picture Fig.~\ref{tcschem} \cite{fp2}, and analytical continuation in determining this critical line thus works.

\section{Conclusions}

The last couple of years have seen an enormous increase in activities concerned with lattice determinations of the QCD phase diagram in all of its interesting regions and limits.
While definite conclusions cannot yet been drawn, there is a lot of progress in refining the methods and studying the systematics.

The longstanding question of the nature of the phase transition in the two flavour theory in the chiral limit is still open. But large volumes are now available, and simulations on $N_t=6$ lattices will be undertaken soon. In combination with now available exact algorithms these will hopefully settle the issue in the near future. 

There are now several groups that are tackling the critical endpoint. 
However, these investigations are
extremely difficult and still carried out over a scatter of theories and parameter values.  Within the R-algorithm, the critical endpoint from two-parameter reweighting is consistent with the shape of the critical surface determined from imaginary chemical potential. However, the R-algorithm in the regime of physical quark mass has been demonstrated to be afflicted by strong stepsize effects, which change the apparent order of the phase transition. Exact algorithms are now being employed successfully, and this source of error will soon be eliminated.
An important qualitative conclusion is that the critical chemical potential of the endpoint is extremely quark mass sensitive. A critical point $\mu_B^c\lsim 400$ MeV requires the physical light quark masses to be less than 5\% larger than the critical values at zero density. While it is quite possible that nature has arranged for this, it is clear that under those circumstances a quantitative determination is going to be a formidable task: any
systemtic error in the current simulations is going to have enormous effects on the location of the critical point. Recall that all calculations reported here are on coarse lattices with $a\sim 0.3$ fm, and in most works quark masses are only fixed in lattice units. Furthermore, finite volume and stepsize effects
have been shown to be larger than several $10\%$. 
Under those circumstances it is still conceivable that there is no critical point and phase transition at all. This means working towards producing results in the thermodynamic and continuum limits will be just as exciting as the first qualitative calculations!

\vspace*{0.5cm}
\noindent
{\bf Acknowledgements:}
I am grateful to Philippe de Forcrand for valuable discussions and comments, help with figures, and a continued enjoyable collaboration.

\end{document}